\documentclass[11pt]{article}
\usepackage[utf8]{inputenc}

\usepackage{verbatim}
\usepackage{caption}
\usepackage{amssymb}
\usepackage{amsmath}
\usepackage{tabularx}
\usepackage{color}
\usepackage{stmaryrd}
\usepackage{multirow}
\usepackage{mathtools}
\usepackage{hyperref}
\usepackage{epsfig,latexsym}

\let\oldequation\equation
\let\oldendequation\endequation

\usepackage{amsmath,amssymb,amsthm,graphicx,verbatim,bm,fancyhdr,bbm,enumerate,color, multirow, tabularx}
\usepackage{hyperref}

\DeclareMathAlphabet{\mathpzc}{OT1}{pzc}{m}{it}  

\theoremstyle{definition}
\theoremstyle{plain}
\newtheorem{theorem}{Theorem}
\newtheorem{proposition}[theorem]{Proposition}
\newtheorem*{proposition*}{Proposition}

\theoremstyle{remark}

\newtheorem*{remark*}{Remark}

\newtheorem*{terminology*}{Terminology}
\newtheorem*{notation*}{Notation}

\newcommand{\bfSigma}{\boldsymbol\Sigma}
\newcommand{\bfalpha}{\boldsymbol\alpha}

\newcommand{\E}{\mathbb{E}}

\let\oldalign\align
\let\oldendalign\endalign

\renewenvironment{align}
  {\linenomathNonumbers\oldalign}
  {\oldendalign\endlinenomath}

\let\oldequation\equation
\let\oldendequation\endequation

\renewenvironment{equation}
  {\linenomathNonumbers\oldequation}
  {\oldendequation\endlinenomath}

\usepackage[mathlines]{lineno}

\title{
Statistical inference
of the rates of cell proliferation and phenotypic switching in cancer
}

\author{Einar Bjarki Gunnarsson$^{1,2,*}$ \and Jasmine Foo$^{2}$ \and Kevin Leder$^1$}

\date{%
    \footnotesize $^1$Department of Industrial and Systems Engineering, University of Minnesota, Twin Cities, MN 55455, USA. \\[2pt]%
    $^2$School of Mathematics, University of Minnesota, Twin Cities, MN 55455, USA. \\
    {* corresponding author} \\[18pt]
}

\renewenvironment{abstract}
 {\small
  \begin{center}
  \bfseries \abstractname\vspace{0pt}\vspace{0pt}
  \end{center}
  \list{}{%
    \setlength{\leftmargin}{13.7mm}
    \setlength{\rightmargin}{\leftmargin}%
  }%
  \item\relax}
 {\endlist}

\begin{document}


\begin{center}
{\bf\large Accepted author manuscript (Journal of Theoretical Biology)}
\end{center}

\vspace*{-12pt}

\begingroup
\let\newpage\relax
\maketitle
\endgroup

\maketitle

\begin{abstract}
Recent evidence suggests that nongenetic (epigenetic) mechanisms play an
important role at all stages of cancer evolution.
In many cancers, these mechanisms have been observed to induce dynamic switching
between two or more cell states, which commonly show differential responses to drug treatments.
To understand how these cancers evolve over time,
and how they respond to treatment,
we need to understand the state-dependent rates of cell proliferation and phenotypic switching.
In this work, we propose a rigorous 
statistical framework for estimating these parameters,
using data from commonly performed cell line experiments,
where phenotypes are sorted and expanded in culture.
The framework explicitly models the stochastic dynamics of cell division, cell death and phenotypic switching,
and it provides likelihood-based confidence intervals
for the model parameters.
The input data can be either the fraction of cells or the number of cells in each state at one or more time points.
Through a combination of theoretical analysis and numerical simulations, we show that when cell fraction data is used,
the rates of switching may be the only parameters that can be estimated accurately.
On the other hand, using cell number data enables accurate estimation of the net division rate for each phenotype,
and it can even enable estimation of the state-dependent rates of cell division and cell death.
We conclude by applying our framework to a publicly available dataset.
\end{abstract}

\vspace*{12pt}

\vspace*{12pt}

\noindent © 2023. This manuscript version is made available under the CC-BY-NC-ND 4.0 license \url{https://creativecommons.org/licenses/by-nc-nd/4.0/}.

\section{Introduction} \label{sec:introduction}

Cancer evolution has long
been understood to be 
a genetic process.
However, recent evidence suggests an equally important role for non-genetic 
forces, 
including 
epigenetic mechanisms and the 
inherent stochasticity in gene transcription and translation
\cite{Brock2009, jones2007epigenomics, brown2014poised, flavahan2017epigenetic,salgia2018genetic,biswas2021drivers}.
These mechanisms are heritable and reversible, 
and they can enable cells to dynamically switch
between two or more 
phenotypic states.
Such switching dynamics have been
observed 
e.g.~in 
lung cancer \cite{sharma2010chromatin,ramirez2016diverse,Engelman2016}, melanoma \cite{roesch2010temporarily,shaffer2017rare,su2017single}, glioblastoma \cite{liau2017adaptive,neftel2019integrative}, leukemia \cite{pisco2013non,knoechel2014epigenetic}, colon cancer \cite{yang2012dynamic,feng2012characterization,geng2014dynamic,wang2014dynamics} and breast cancer \cite{gupta2011stochastic,goldman2015temporally,jordan2016her2,bhatia2019interrogation}.
The different phenotypes
commonly show differential responses to drug treatments,
which enhances 
the adaptability of the cancer
under treatment
and significantly 
increases the probability of treatment resistance
\cite{gunnarsson2020understanding}.

Unraveling how the cancer-specific rates of cell division, cell death 
and phenotypic switching shape tumor evolution over time
is crucial to furthering our understanding of the disease
and to informing new 
treatment strategies.
For example, in a two-phenotype cancer where one type is drug-sensitive
and the other is drug-tolerant,
the change in phenotypic proportions
during the initial stages of treatment
can be explained by a combination of sensitive cells dying,
drug tolerant cells proliferating,
and cells switching between sensitivity and tolerance.
Disentangling the relative rates at which these events occur
can help us to better understand how resistance arises,
how it evolves over time,
and how best to combat it \cite{gunnarsson2020understanding}.

\begin{figure}[!t]
    \centering
    \includegraphics[scale=0.5]{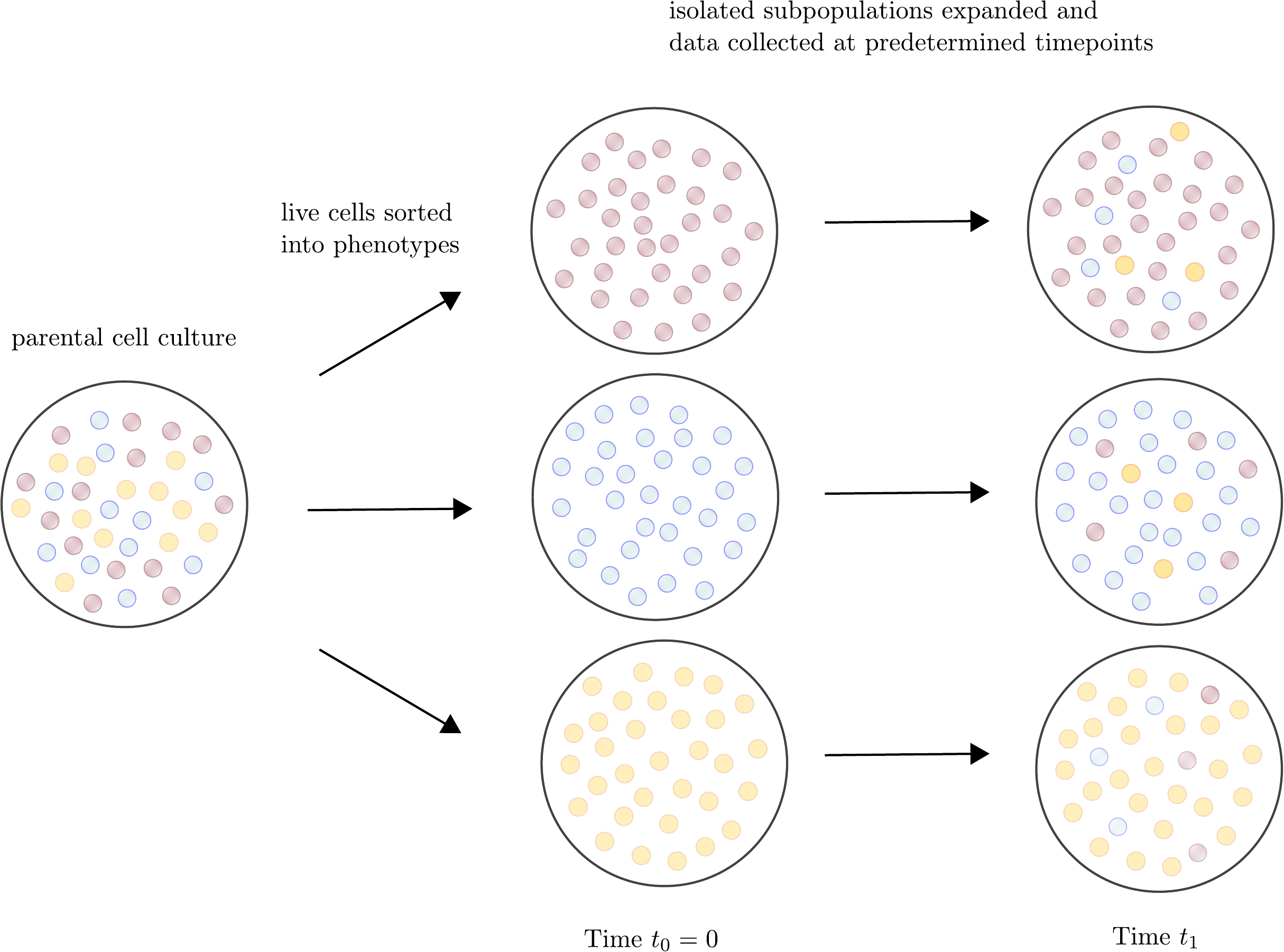}
    \caption[ The 
    dynamics of phenotypic switching are commonly
    interrogated by sorting live cells into isolated phenotypic subpopulations
    and expanding the subpopulations in culture.]{
    The 
    dynamics of phenotypic switching are commonly
    interrogated by sorting live cells into 
    isolated phenotypic subpopulations
    and expanding these subpopulations in culture \cite{gupta2011stochastic,yang2012dynamic,geng2014dynamic,wang2014dynamics,jordan2016her2,bhatia2019interrogation,neftel2019integrative}.
  By tracking the evolution of phenotypic proportions over time
  and applying mathematical models of phenotypic switching,
  it becomes possible to estimate the quantitative parameters of the process \cite{gupta2011stochastic,buder2017celltrans,jagannathan2020transcompp,li2022mathematical,zhou2014nonequilibrium,goldman2015temporally,su2017single,devaraj2019morphological}.
  }
    
    \label{fig:FACSexplanation}
\end{figure}

Our current quantitative understanding of the rates of cell proliferation and 
phenotypic switching in cancer
is largely derived from cell line experiments.
In these experiments, live cells are commonly sorted into phenotypes, 
e.g.~based on gene expression profiles or cell morphologies,
isolated subpopulations are expanded in culture,
and phenotypic proportions
are tracked over time (Fig.~\ref{fig:FACSexplanation}). 
These isolated subpopulations have 
been observed to 
give rise to all other phenotypes over time,
with proportions between types eventually converging to the 
constant proportions 
observed in the parental population
\cite{gupta2011stochastic,yang2012dynamic,geng2014dynamic,wang2014dynamics,jordan2016her2,bhatia2019interrogation,neftel2019integrative}.

To explain this behavior, simple mathematical models of
phenotypic switching have been proposed,
and these models have 
been used to 
estimate the rates at which
cells switch between states \cite{gupta2011stochastic,buder2017celltrans,jagannathan2020transcompp,li2022mathematical,zhou2014nonequilibrium,goldman2015temporally,su2017single,devaraj2019morphological}.
These works
are reviewed 
in Section
\ref{app:litreview} below.
Previous estimation methods have been deterministic in nature,
and they have generally 
derived their estimates from data
on the fraction of cells in each state
at each time point.
If 
the total size of the cell population
is measured at the same time points,
as e.g.~in \cite{devaraj2019morphological},
one obtains data on the number of cells in each state at each time point.
We will show that when cell fraction data is used,
the rates of phenotypic switching may be the only parameters
that can be estimated accurately.
In contrast, using cell number data
enables accurate
estimation of the net cell division rate
for each phenotype,
and
it can even enable estimation of the 
state-dependent
rates of cell division and cell death.
Understanding 
how growth rates vary between types 
is as important as understanding
the rates of phenotypic switching,
especially in the context of treatment response.
Not only do
the growth rates 
influence the phenotypic composition of the population,
they also control the evolution of the tumor burden over time.

Our goal in this work is to develop a statistically rigorous framework for estimating the rates of cell proliferation and phenotypic switching in cancer.
In contrast to previous approaches, 
our framework 
explicitly models the stochastic dynamics of
cell division, cell death and phenotypic switching,
it provides likelihood-based confidence intervals for the model parameters,
and it enables estimation
both from cell fraction and cell number data.
We also use 
our framework to analyze
the identifiability of model parameters and how it 
depends on the input data.
This important topic has not been addressed by previous works.

The rest of the paper is organized as follows.
In Section \ref{app:litreview}, 
we review prior estimation methods.
In Section \ref{sec:model}, we introduce our stochastic model of 
cell division, cell death and phenotypic switching.
In Section \ref{sec:experimentsanddata}, we 
state our assumptions 
on the cell line experiments conducted
and 
the data collected.
In Sections \ref{sec:estimationnum} and \ref{sec:estimationfrac},
we propose statistical models for cell number and cell fraction data, respectively,
and describe how parameter estimates and confidence intervals are computed.
In Section \ref{sec:identifiability},
we present theoretical analysis of 
the identifiability of parameters
under each model.
In Section \ref{sec:numerical}, we conduct numerical experiments for the case of two phenotypes,
and in Section \ref{sec:realdata}, we apply our framework to a publicly available dataset.
We conclude by discussing limitations
of the framework as well as avenues for improvement (Section \ref{sec:discussion}).
For simplicity, the development of the estimation framework in the main text is focused on the case of experiments started by isolated subpopulations.
General starting conditions are treated in full detail in the appendices.

\section{Review of prior estimation methods} \label{app:litreview}

At the single-cell-level, phenotypic switching has commonly been modeled by a discrete-time Markov chain with $K \geq 2$ states, where $K$ is the number of phenotypes.
In each time step, a cell in state $j$ transitions to state $k \neq j$ with probability $p_{jk}$, 
and it remains in state $j$ with probability $p_{jj} = 1-\sum_{k \neq j} p_{jk}$.
The transition probabilities are collected into the $K \times K$ {\em transition matrix} ${\bf P} = (p_{jk})$.
The evolution of the Markov chain is determined by ${\bf P}$ and the {\em initial distribution}
${\bf q} = (q_1,\ldots,q_K)$,
where 
$q_j$
is the probability that a cell starts in state $j$.
If we let ${\bf q}^{(\ell)}$ denote the cell state distribution after $\ell \geq 1$ time steps,
then ${\bf q}^{(\ell)} = {\bf q}{\bf P}^{\ell}$.

Say we conduct $K$ cell line experiments starting with $N$ cells in each experiment
and known initial cell state distributions ${\bf q}_1,\ldots,{\bf q}_K$.
The initial distributions are collected into a $K \times K$ matrix ${\bf Q}$,
where ${\bf q}_i$ is the $i$-th row vector.
Each experiment is run for $\ell \geq 1$ time steps, at which point the fraction of cells in each state is recorded.
Let $f_{ij}^{(\ell)}$ be the observed fraction of cells in state $j$ under the $i$-th initial condition.
The observations at the $\ell$-th time step under the $i$-th initial condition are collected into a vector ${\bf f}_i^{(\ell)} = \big(f_{i1}^{(\ell)},\ldots,f_{iK}^{(\ell)}\big)$,
and all observations at the $\ell$-th time step are collected 
into a $K \times K$ matrix ${\bf F}^{(\ell)} = \big(f_{ij}^{(\ell)}\big)$.
If there are multiple replicates $r=1,\ldots,R$,
we let 
${\bf F}^{(\ell),r}$ denote the data from the $r$-th replicate.

Now assume that the starting population $N$ is large,
that there is no cell division or cell death,
and that each cell switches between states
according to the above Markov model.
In this case, by the strong law of large numbers,
the model-predicted distribution between cell states ${\bf Q}{\bf P}^\ell$
after $\ell$ time steps
can be approximated by the experimentally observed cell-state fractions ${\bf F}^{(\ell)}$.
If we simply equate these two matrices,
we can obtain an estimate ${\bf P}_\ell$ of the transition matrix 
${\bf P}$ by inverting the matrix ${\bf Q}$ of initial distributions and taking an $\ell$-th matrix root,
${\bf P}_\ell = \big({\bf Q}^{-1}{\bf F}^{(\ell)}\big)^{1/\ell}$.
Here, we assume that ${\bf Q}$ is invertible,
which is e.g.~the case when experiments are started with isolated subpopulations.

This simple estimation idea was applied by Gupta et al.~\cite{gupta2011stochastic}
to investigate phenotypic switching between stem-like, basal and luminal cell states in breast cancer,
using data from a single time point.
A multiple-time-point extension has since been implemented
in the R package CellTrans \cite{buder2017celltrans}.
Say that cell state fractions are experimentally observed at time steps
$m_1,\ldots,m_L$ for $L \geq 1$.
CellTrans first computes an estimate ${\bf P}_{m_\ell}$ of the transition matrix for each time step as above,
and then returns a final estimate as the average across time steps:
\begin{align} \label{eq:CellTransestimate}
    \textstyle \widehat{\bf P} := (1/L) \sum_{\ell=1}^L {\bf P}_{m_\ell}.
\end{align}
CellTrans also involves a regularization step
to ensure that
$\widehat{\bf P}$ is stochastic.
CellTrans is used on publicly available datasets in \cite{buder2017celltrans}
and it has been applied more recently in
\cite{dirkse2019stem,vipparthi2022emergence,chedere2021multi}.

Cell populations in culture typically change in size over time.
If all phenotypes grow at the same rate,
and cell growth occurs deterministically at the end of each time step,
the constant-sized Markov model can be used to
describe the evolution of cell state fractions.
Both Gupta et al.~\cite{gupta2011stochastic} and 
Su et al.~\cite{su2017single} have applied an augmented version of 
the Markov model
intended to capture 
proliferation differences between types.
In the augmented model, during a single time step,
each type-$j$ cell first grows deterministically to a population of size $\Lambda_{jj}$, and a fraction $p_{jk}$ of cells then switch to type-$k$.
The growth factors $\Lambda_{jj}$ are collected into a diagonal proliferation matrix $\boldsymbol\Lambda$, 
and the multiple $\boldsymbol\Lambda {\bf P}$, after being normalized to produce cell fractions as opposed to cell numbers, is used to predict the distribution between cell states.
In both Gupta et al.~\cite{gupta2011stochastic} and Su et al.~\cite{su2017single},
the  matrix $\boldsymbol\Lambda$ is found by
randomly sampling candidate parameter values
and selecting the values that best fit the experimental data.

TRANSCOMPP \cite{jagannathan2020transcompp} is a more systematic version of 
the aforementioned method.
In TRANSCOMPP,
the diagonal proliferation matrix $\boldsymbol\Lambda$
and the transition matrix ${\bf P}$
are estimated by minimizing the sum of squared errors
between the model prediction and the data,
\begin{align}
\label{eq:TRANSCOMPP}
    \textstyle \min_{\boldsymbol\Lambda, {\bf P}} \sum_{i=1}^I \sum_{\ell=1}^L \sum_{r=1}^R \Big\|{\bf f}_i^{({m_\ell}),r}-\big({\bf q}_i\big(\boldsymbol\Lambda{\bf P}\big)^{m_\ell}{\bf 1}^T\big)^{-1} {\bf q}_i \big(\boldsymbol\Lambda{\bf P}\big)^{{m_\ell}}\Big\|^2.
\end{align}
Note that this problem only determines the growth factors relative to one another, $\Lambda_{jj}/\Lambda_{11}$ for $j=2,\ldots,K$.
TRANSCOMPP is implemented in MATLAB, 
and it includes a stochastic resampling procedure for estimating
the distributions of the transition probability estimates.
The stochastic resampling is performed on single-cell measurements of cell phenotypes, if available, or on data generated from a user-defined distribution of cell state fractions.

In modeling switching between HER2$+$ and HER2$-$ states in breast cancer,
Li and Thirumalai \cite{li2022mathematical} employ
a deterministic continuous-time model.
Their model assumes symmetric and asymmetric cell divisions,
which through reparametrization leads to the same dynamics as symmetric cell divisions and switching between types.
Li and Thirumalai assume equal rates of
asymmetric division for the two types (or equivalently, equal rates of switching between types),
and they show that if experiments are started with isolated subpopulations,
the slopes of the cell fraction trajectories at time 0
can be used to estimate these rates.
They also show that the equilibrium proportion between types
can be used to estimate the difference in symmetric division rate between the two types.
The proportion between phenotypes in the parental population
is used as an estimate of the equilibrium proportion.
We have made use of these insights in
our identifiability analysis in Section \ref{sec:identifiabilitycellfraction} 
below.

Finally, in their 
investigation of epithelial to mesenchymal transition in breast cancer, Devaraj and Bose \cite{devaraj2019morphological,devaraj2020mathematics}
employ a discrete-time model
where cells divide, die and switch between types.
Their model includes a separate state for dead cells
to facilitate estimation of death rates and well as division rates.
We have used the same idea in Section \ref{sec:birthrates} below
to improve the identifiability of birth and death rates under our framework.
Their model furthermore assumes that the rates of birth, death and switching are time-dependent.
Devaraj and Bose derive difference equations
for the change in the number of cells in each state between time points.
They then propose a multi-objective optimization problem
to estimate the model parameters from data on 
cell state fractions 
and the total number of alive and dead cells at each time point.
Their parameter fitting procedure
minimizes the least squares error between
the model predictions and the data
across the different time points,
while ensuring that parameters do not vary too drastically
between time periods.

\section{Models and methods}

In this section, we propose statistical models for cell number and cell fraction data,
which are based on a multitype branching process model of the cell population dynamics \cite{athreya2004branching}.
To simplify the discussion, we will 
focus on the case where 
all experiments are started from isolated subpopulations of cells.
We emphasize however that the estimation framework can be applied to any set of starting conditions,
as is outlined in more detail 
in Appendix \ref{app:estimationdeets}.

\subsection{Multitype branching process model}\label{sec:model}

\subsubsection{Model definition and model parameters} \label{sec:modelparameters}

\begin{figure}
    \centering
    \includegraphics[scale=0.9]{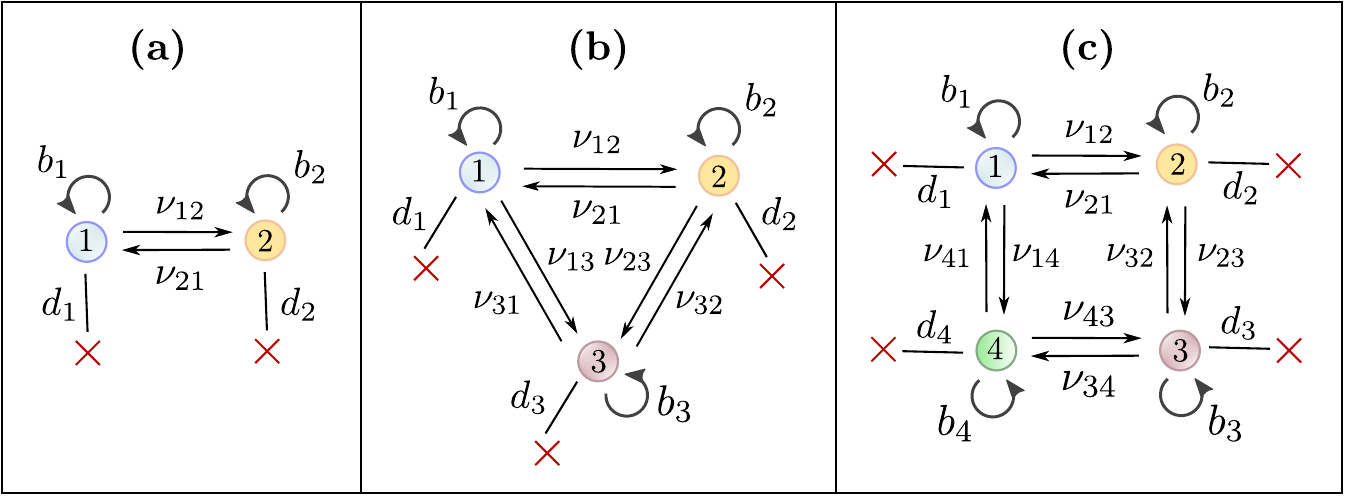}
    \caption[The multitype branching process model captures a variety of switching dynamics.]{
    The multitype branching process model captures a variety of switching dynamics previously observed in the literature.
    {\bf (a)} A two-type model captures e.g.~the dynamics between HER2+ and HER2$-$ cell states in Brx-82 and Brx-142 breast cancer cells \cite{jordan2016her2}.
    {\bf (b)} A three-type model captures e.g.~the dynamics between stem-like, basal and luminal cell states in SUM149 and SUM159 breast cancer cells \cite{gupta2011stochastic}.
    {\bf (c)} A four-type model captures e.g.~the dynamics between CD24$^{\rm Low}$/ALDH$^{\rm High}$, CD24$^{\rm Low}$/ALDH$^{\rm Low}$, CD24$^{\rm High}$/ALDH$^{\rm High}$ and CD24$^{\rm Hich}$/ALDH$^{\rm Low}$ cell states in GBC02, SCC029B and SCC070
    oral cancer cells \cite{vipparthi2022emergence}.
    }
    \label{fig:modelstructures}
\end{figure}

To model the cell population dynamics,
we employ 
a multitype 
branching process model in continuous time, 
with $K \geq 2$ types \cite{athreya2004branching}.
In the model, a type-$j$ cell divides into two cells at rate $b_j \geq 0$, it dies at rate $d_j \geq 0$, and it switches to type-$k$ at rate $\nu_{jk} \geq 0$ for $k \neq j$,
independently of all other cells.
This means that in an infinitesimally short time interval of length $\Delta t>0$,
a type-$j$ cell divides with probability $b_j\Delta t$, it dies with probability $d_j\Delta t$, and it switches to type $k$ with probability $\nu_{jk} \Delta t$.
The multitype branching process model captures a variety of switching dynamics previously observed in the literature (Fig.~\ref{fig:modelstructures}).

We allow $\nu_{jk}=0$ for some $j$ and $k$, which means that a type-$j$ cell is not able to switch directly to type-$k$.
However, in our exposition, we assume  
that the 
model is {\em irreducible},
in that each cell type is accessible from any other cell type,
possibly through intermediate types.
In mathematical terms, this means that
for each $j,k=1,\ldots,K$ with $k \neq j$, 
there exist $r \geq 0$ integers $m_1,\ldots,m_r \in \{1,\ldots,K\}$ so that $\nu_{jm_1}\nu_{m_1m_2}\cdots\nu_{m_rk}>0$.
Our estimation framework can also be applied to reducible switching models,
as we discuss in 
Appendix \ref{sec:extensions} below.

For $j=1,\ldots,K$, we define $\lambda_j := b_j-d_j$ as the net birth rate of a type-$j$ cell.
We collect the growth parameters into $1 \times K$ vectors ${\bf b} = (b_1,\ldots,b_K)$, ${\bf d}=(d_1,\ldots,d_K)$ and $\boldsymbol\lambda = (\lambda_1,\ldots,\lambda_K)$.
We also define ${\boldsymbol\lambda}^{[-j]} := \boldsymbol\lambda - \lambda_j{\bf 1}$ as the vector of net birth rates relative to $\lambda_j$, with $\lambda^{[-j]}_k = \lambda_k-\lambda_j$ for $k \neq j$ and $\lambda^{[-j]}_j = 0$.
We 
finally define 
the $K \times K$ matrix ${\bf A}$ with $a_{jj} := \lambda_j-\sum_{k \neq j} \nu_{jk}$ for $j=1,\ldots,K$ and $a_{jk} := \nu_{jk}$ for $k \neq j$ as the {\em infinitesimal generator} of the model, where
$a_{jk}$ is the net rate at which a cell of type $j$ produces a cell of type $k$.

\subsubsection{Random processes and their moments} \label{sec:randomproc}

If the branching process is started by $N$ cells of type-$j$, the state of the process at time $t \geq 0$ is encoded in the $1 \times K$ random vector of cell numbers ${\bf Z}^{(j)}(t;N) = \big(Z^{(j)}_1(t;N),\ldots,Z^{(j)}_K(t;N)\big)$.
On the event $\big\{\sum_{k=1}^K Z_k^{(j)}(t;N) \neq 0\big\}$
that the cell population is still alive at time $t$,
we 
let $\boldsymbol\Delta^{\!(j)}(t;N)$ denote the corresponding random vector of cell fractions, with
\[
\textstyle \Delta_i^{\!(j)}(t;N) := Z_i^{(j)}(t;N)/\big(\sum_{k=1}^K Z_k^{(j)}(t;N)\big), \quad i=1,\ldots,K.
\]
If the process is started by a single type-$j$ cell,
we write ${\bf Z}^{(j)}(t) := {\bf Z}^{(j)}(t;1)$,
and we define the associated mean vector and covariance matrix by
\begin{align} \label{eq:mSigma}
\begin{split}
        & {\bf m}^{(j)}(t) := \E\big[{\bf Z}^{(j)}(t)\big], \\
    & {\boldsymbol\Sigma}^{(j)}(t) := \E\big[\big({\bf Z}^{(j)}(t)-{\bf m}^{(j)}(t)\big)^T\big({\bf Z}^{(j)}(t)-{\bf m}^{(j)}(t)\big)\big], \quad t \geq 0.
\end{split}
\end{align}
We also define the $K \times K$ matrix ${\bf M}(t)$ with row vectors
${\bf m}^{(j)}(t)$ as the {\em mean matrix} for the process at time $t$.
It can be shown that ${\bf M}(t)$ is given by the matrix exponential $\exp(t{\bf A}) := \sum_{k=0}^\infty \frac{t^k}{k!} {\bf A}^k$  \cite{athreya2004branching}.
Note that ${\bf A}$ and ${\bf M}(t)$ depend on the birth rates ${\bf b}$ and the death rates ${\bf d}$ only through the net birth rates $\boldsymbol\lambda$.

                 \begin{table}[t]
    \centering
    \renewcommand{\arraystretch}{1.1}
    \begin{tabular}{|l|l|l|}
    \hline 
    Symbol&Dimension&Description \\
    \hline
            $K$ & 1 & Number of types \\
    $b_j$ & 1&Division rate of type-$j$ cells  \\
    $d_j$ & 1&Death rate of type-$j$ cells  \\
    $\nu_{jk}$ & 1&Rate of switching from type-$j$ to type-$k$  \\ 
    $\lambda_j$ & 1& Net birth rate of type-$j$ cells, $\lambda_j=b_j-d_j$  \\
        $\lambda_k^{[-j]}$ & 1 & Net birth rate relative to $\lambda_j$, $\lambda_k^{[-j]} = \lambda_k-\lambda_j$ \\
        \hline
    ${\bf A}$ & $K\times K$ &Infinitesimal generator of the branching process model \\
    ${\bf M}(t)$ & $K\times K$ & Mean matrix at time $t$, ${\bf M}(t) = \exp(t{\bf A})$ \\
    $\overline{\boldsymbol\gamma}$ & $1 \times K$ & Equilibrium proportions between cell types \\
    \hline
    ${\bf Z}^{(j)}(t;N)$ & $1\times K$ & Vector of cell numbers at time $t$, started by $N$ type-$j$ cells \\   
        ${\bf Z}^{(j)}(t)$ & $1\times K$ &  ${\bf Z}^{(j)}(t) :=  {\bf Z}^{(j)}(t;1)$ \\    
                        ${\bf m}^{(j)}(t)$ & $1\times K$ & ${\bf m}^{(j)}(t) := \E\big[{\bf Z}^{(j)}(t)\big] = {\bf e}_j{\bf M}(t)$ \\
                            ${\boldsymbol\Sigma}^{(j)}(t)$ & $K\times K$ & ${\boldsymbol\Sigma}^{(j)}(t) := \E\big[\big({\bf Z}^{(j)}(t)-{\bf m}^{(j)}(t)\big)^T\big({\bf Z}^{(j)}(t)-{\bf m}^{(j)}(t)\big)\big]$ \\
    \hline
    ${\boldsymbol\Delta}^{\!(j)}(t;N)$ & $1 \times K$ & Vector of cell fractions at time $t$, started by $N$ type-$j$ cells \\
        ${\boldsymbol\Delta}^{\!(j)}(t)$ & $1 \times K$ & ${\boldsymbol\Delta}^{(j)}(t) := {\boldsymbol\Delta}^{\!(j)}(t;1)$  \\
        ${\bf p}^{(j)}(t)$ & $1 \times K$ & ${\bf p}^{(j)}(t) 
:= \big({\bf m}^{(j)} {\bf 1}^T\big)^{-1} {\bf m}^{(j)}(t)$ \\
    ${\bf S}^{(j)}(t)$ & $K \times K$& ${\bf S}^{(j)}(t) := \big({\bf m}^{(j)}(t){\bf 1}^T\big)^{-2}\big({\bf I} - {\bf 1}^T {\bf p}^{(j)}(t)\big)^T \, {\boldsymbol\Sigma}^{(j)}(t)  \, \big({\bf I} - {\bf 1}^T {\bf p}^{(j)}(t)\big)$ \\
    \hline
    \end{tabular}
    \caption{
    Notation defined in Section \ref{sec:model} and Section \ref{sec:estimationfrac}.
    }
    \label{table:notationmodel}
\end{table}

\subsubsection{Long-run behavior} \label{sec:longrunbehavior}

In the branching process model
with irreducible switching dynamics,
all subpopulations eventually grow at the same exponential rate $\sigma$.
This applies both to individual trajectories of the model (when the population does not go extinct)
and its mean behavior.
In mathematical terms, 
if the process is started by a single type-$j$ cell,
there exists a real number $\sigma$,
positive $1 \times K$ vectors 
${\boldsymbol\beta} = (\beta_1,\ldots,\beta_K)$
and ${\boldsymbol\gamma} = (\gamma_1,\ldots,\gamma_K)$,
and a nonnegative random variable $W$ with mean $\E[W] = \beta_j$, so that
\begin{align} \label{eq:almostsureconvergence}
        \textstyle \lim_{t \to \infty} e^{-\sigma t} {\bf Z}^{(j)}(t) = W\boldsymbol\gamma, 
\quad \text{almost surely},
\end{align}
and
\begin{align} \label{eq:Meanmatrixconvergence}
        & \textstyle \lim_{t \to \infty} e^{-\sigma t} {\bf m}^{(j)}(t) = \beta_j\boldsymbol\gamma.
\end{align}
See e.g.~Sections V.7.1-V.7.4 and Theorem 2 in Section V.7.5 of \cite{athreya2004branching}.
In other words, the number of type-$k$ cells at time $t$ is approximately $W \gamma_ke^{\sigma t}$ almost surely when $t$ is large, and the mean number of type-$k$ cells is approximately $\beta_j \gamma_k e^{\sigma t}$.
It follows that
if we define 
\begin{align} \label{eq:vbardef}
\textstyle \overline{\gamma}_k := \gamma_k/\big(\sum_{m=1}^K \gamma_m\big), \quad k=1,\ldots,K,    
\end{align}
then given that the population does not go extinct,
$\overline{\gamma}_k$ is the long-run proportion of type-$k$ cells in the population,
independently of the initial condition.
Thus, in the long run, 
cell proportions tend towards an equilibrium distribution given by $\overline{\boldsymbol\gamma}$,
which 
is consistent
with the experimental observations discussed in the introduction.

\subsection{Experimental assumptions and notation for experimental data} \label{sec:experimentsanddata}

In the development of our estimation framework,
we assume that each experiment returns measurements from a single time point only, 
meaning that the experimental sample is discarded once measurements are taken
(endpoint data).
In this case, techniques such as flow cytometry or fluorescence-activated cell sorting (FACS)
can be used to identify phenotypes at the experimental endpoints.
Sometimes, the data collected is sequential,
meaning that a single experiment returns measurements from multiple time points.
This can for example be the case when phenotypes are tagged with fluorescent dyes
and tracked over time using time-lapse microscopy (live-cell imaging) \cite{bintu2016dynamics,nam2022dynamic}.
In Section \ref{sec:endpointvssequential}, we show that our endpoint-data statistical framework can also yield
reasonable estimates for sequential data.
In Appendix \ref{app:estimationcellnum}, we discuss what would be required to rigorously extend the framework to sequential data.

In the main text, we assume that each experiment is started by an isolated subpopulation,
and we let $N_j$ be the number of starting cells for the experiment started only by type-$j$ cells.
We assume that $N_j$ is large,
which is generally the case for the 
experiments discussed in the introduction (Section \ref{sec:introduction}).
Furthermore let 
$0 < t_1 < t_2 < \cdots < t_{L}$ with $L \geq 1$ denote the time points at which data is collected, and 
let $R \geq 1$ be the number of experimental replicates performed.
The data collected in each experiment is either a vector ${\bf n}_{j,\ell,r} = (n_{j,\ell,r,1},\ldots,n_{j,\ell,r,K})$ of cell numbers or a vector ${\bf f}_{j,\ell,r} = (f_{j,\ell,r,1},\ldots,f_{j,\ell,r,K})$ of cell fractions.
Here, $n_{j,\ell,r,k}$ is the number of type-$k$ cells in the $r$-th replicate of the experiment started only by type-$j$ cells 
and ended at the $\ell$-th timepoint,
and $f_{j,\ell,r,k}$ is the corresponding cell fraction.

\subsection{Estimation for cell number data} \label{sec:estimationnum}

Our statistical framework for cell number data is rooted in a central limit theorem for the vector ${\bf Z}^{(j)}(t;N)$ of cell numbers at time $t$.
More precisely, by decomposing the branching process $\big({\bf Z}^{(j)}(s;N)\big)_{s \geq 0}$ into i.i.d.~processes started by single type-$j$ cells, we can show that as $N \to \infty$,
\begin{align} \label{eq:CLTnum}
        & N^{-1/2}\big({\bf Z}^{(j)}(t;N)-N{\bf m}^{(j)}(t)\big) \stackrel{d}{\to} {\cal N}\big({\bf 0},\bfSigma^{(j)}(t)\big).
    \end{align}
    The details are provided in Appendix \ref{app:CLT1}, where we also show that the
covariance matrix ${\boldsymbol\Sigma}^{(j)}(t)$ 
is given by
        \begin{align} \label{eq:covmatrixexpression}
    \begin{split}
                \textstyle {\boldsymbol\Sigma}^{(j)}(t) &= \textstyle  2 \int_0^t ({\bf M}(t-\tau))^T {\rm diag}\big({\bf b} \odot {\bf m}^{(j)}(\tau)\big) ({\bf M}(t-\tau))d\tau \\
        &\quad \textstyle +{\rm diag}\big({\bf m}^{(j)}(t)\big)- ({\bf m}^{(j)}(t))^T{\bf m}^{(j)}(t).
    \end{split}
    \end{align}
    When 
    the starting cell number $N$ is large,
the central limit theorem \eqref{eq:CLTnum} allows us to approximate the distribution of ${\bf Z}^{(j)}(t;N)$ by a multivariate normal distribution as follows:
\begin{align} \label{eq:normalapprox}
    {\bf Z}^{(j)}(t;N) \approx N {\bf m}^{(j)}(t) + {\cal N}\big({\bf 0}, N {\boldsymbol\Sigma}^{(j)}(t)\big).
\end{align}
Based on this approximation, we propose the following statistical model for the 
experimental data ${\bf n}_{j,\ell,r}$:
\begin{align} \label{eq:statisticalmodelnum}
    {\bf n}_{j,\ell,r} 
    \; \sim \;
    \underbrace{N_j {\bf m}^{(j)}(t_\ell)}_{\substack{{\rm mean} \\{\rm behavior}}} \;+\; \underbrace{{\cal N}\big({\bf 0}, N_j {\boldsymbol\Sigma}^{(j)}(t_\ell)\big)}_{\substack{{\rm variability \; in} \\ {\rm population \; dynamics}}} \;+\; \underbrace{{\cal N}\big({\bf 0}, {\bf E}_{j,\ell}^{\rm num}\big)}_{\substack{{\rm measurement} \\{\rm error}}}.
\end{align}
The first two terms capture the mean and variance of the 
branching process model dynamics,
while the final term captures experimental measurement error,
which is independent of the branching process.
We assume that the $K \times K$ covariance matrix ${\bf E}_{j,\ell}^{\rm num}$ associated with  measurement error can be written as a function 
of 
the branching process model parameters 
and additional error
parameters
$\boldsymbol\omega_{\rm num} = (\omega_{1},\ldots,\omega_{M_{\rm num}})$
for some $M_{\rm num} \geq 0$.
A simple example is
${\bf E}_{j,\ell}^{\rm num}= \omega^2{\bf I}$ for some $\omega>0$,
where the measurement error is assumed to be of equal magnitude for all data points,
and to be uncorrelated between cell types.
Another simple example is
${\bf E}_{j,\ell}^{\rm num} = \omega^2 \big({\rm diag}\big(N_j{\bf m}^{(j)}(t_\ell)\big)\big)^2$,
where the measurement error is assumed to
scale with mean experimental outcomes.

To compute parameter estimates from the statistical model \eqref{eq:statisticalmodelnum},
we use a maximum likelihood approach, due to its simplicity and desirable large-sample properties like consistency and asymptotic efficiency \cite{casella2021statistical}.
More precisely, 
the statistical model \eqref{eq:statisticalmodelnum}
is used to derive 
a likelihood function,
which is the probability of observing the experimental data as a function of the model parameters,
and point estimates for the parameters are computed
by maximizing the likelihood function.
We also derive a likelihood-based confidence interval for each model parameter $\theta$,
which is obtained by inverting the likelihood-ratio test for the given parameter, i.e.~collecting all values $\theta_0$ for which the null hypothesis $\theta = \theta_0$ is accepted under the likelihood-ratio test \cite{neale1997use,fischer2021robust,borisov2020confidence,venzon1988method,raue2009structural}.
The confidence interval is determined by the profile log-likelihood for $\theta$, as is further discussed in Appendix \ref{app:estimationcellnum}.

\subsection{Estimation for cell fraction data} \label{sec:estimationfrac}

For cell fraction data, we propose a similar maximum
likelihood estimation framework,
rooted in a central limit theorem for the vector ${\boldsymbol\Delta}^{{(j)}}(t;N)$ of cell fractions at time $t$.
To state the central limit theorem, we define the $1 \times K$ vector ${\bf p}^{(j)}(t)$ and the $K \times K$ matrix ${\bf S}^{(j)}(t)$ by
\begin{align} \label{eq:QalphaSalphadef}
\begin{split}
        & {\bf p}^{(j)}(t) :=  \big({\bf m}^{(j)} {\bf 1}^T\big)^{-1} {\bf m}^{(j)}(t), \\
    & {\bf S}^{(j)}(t) := \big({\bf m}^{(j)}(t){\bf 1}^T\big)^{-2}\big({\bf I} - {\bf 1}^T {\bf p}^{(j)}(t)\big)^T \, {\boldsymbol\Sigma}^{(j)}(t)  \, \big({\bf I} - {\bf 1}^T {\bf p}^{(j)}(t)\big).
\end{split}
\end{align}
Using arguments of Yakovlev and Yanev \cite{yakovlev2009relative}, we can show that as $N \to \infty$,
    \begin{align} \label{eq:CLTfrac}
        & N^{1/2}\big(\boldsymbol\Delta^{\!(j)}(t;N)-{\bf p}^{(j)}(t)\big) \stackrel{d}{\to} {\cal N}\big({\bf 0},{\bf S}^{(j)}(t)\big).
    \end{align}
The details are provided in Appendix \ref{app:CLT2}, where we also show that the mean function ${\bf p}^{(j)}(t)$ can be written solely as a function of the switching rates $(\nu_{ik})_{k \neq i}$ and the relative net birth rates $\boldsymbol\lambda^{[-1]}$.
The choice of type-1 as a reference phenotype is arbitrary.
Based on the central limit theorem \eqref{eq:CLTfrac},
we propose the following statistical model
for the experimental data ${\bf f}_{j,\ell,r}$:
\begin{align} \label{eq:statisticalmodelfrac}
    {\bf f}_{j,\ell,r} \; {\sim} \; {\bf p}^{(j)}(t_\ell) + {\cal N}\big({\bf 0},N_j^{-1}{\bf S}^{(j)}(t_\ell)\big) + {\cal N}\big({\bf 0}, {\bf E}_{j,\ell}^{\rm frac}\big).
\end{align}
As for cell number data, we assume 
that the $K \times K$ covariance matrix ${\bf E}_{j,\ell}^{\rm frac}$ associated with measurement error can be written as a function 
of 
the branching process model parameters
and additional error parameters $\boldsymbol\omega_{\rm frac} = (\omega_{1},\ldots,\omega_{M_{\rm frac}})$
for some $M_{\rm frac} \geq 0$.

Note that in the statistical model \eqref{eq:statisticalmodelfrac}, the variability term $N_j^{-1}{\bf S}^{(j)}(t_\ell)$ decreases with the initial population size $N_j$.
Thus, if a large $N_j$ is coupled with a large measurement error, 
the third term in \eqref{eq:statisticalmodelfrac} will dominate the second term.
When applying the framework to real cell fraction datasets, this can potentially allow us to simplify the model in \eqref{eq:statisticalmodelfrac} so that it only includes the first and third term:
\begin{align} \label{eq:statisticalmodelfracsimple}
    {\bf f}_{j,\ell,r} \; {\sim} \; {\bf p}^{(j)}(t_\ell) + {\cal N}\big({\bf 0}, {\bf E}_{j,\ell}^{\rm frac}\big).
\end{align}
We discuss this point further in Section \ref{sec:realdata} and the discussion section (Section \ref{sec:discussion}).

As for cell number data, from the statistical model \eqref{eq:statisticalmodelfrac} (and the simpler version \eqref{eq:statisticalmodelfracsimple}),
it is straightforward to derive a likelihood function, maximum likelihood estimates 
and likelihood-based confidence intervals, as is discussed in more detail in Appendix \ref{app:estimationcellfrac}.

\section{Results}

\subsection{Structural identifiability analysis} \label{sec:identifiability}

We begin by analyzing the structural identifiability of the statistical models \eqref{eq:statisticalmodelnum} and \eqref{eq:statisticalmodelfrac}.
Informally, structural identifiability refers to whether a parameter can be estimated accurately given an infinite amount of noise-free data.
More precisely, a parameter is structurally identifiable if complete knowledge of the model distribution uniquely determines the value of the parameter, in the absence of any measurement noise \cite{rothenberg1971identification,browning2020identifiability}.

To demonstrate the structural identifiability of a parameter, it is sufficient to show that knowledge of the statistical moments of the model distribution implies knowledge of the parameter.
By considering the moments,
we can adopt techniques from systems biology used for the analysis of
deterministic models based on ordinary differential equations \cite{chis2011structural}.
In particular,
we will assume that we know the behavior of the mean functions ${\bf m}^{(j)}(t)$ and ${\bf p}^{(j)}(t)$
and the covariance functions ${\boldsymbol\Sigma}^{(j)}(t)$ and ${\bf S}^{(j)}(t)$
close to time 0 (more precisely, their derivatives at 0),
and we will analyze to what extent the model parameters can be extracted from this information.
In other words, we are interested in the following question:
If we conduct experiments started from isolated subpopulations,
and 
perfect observations are made of
the first two statistical moments of the model close to time 0,
can we identify the model parameters?

This analysis serves two purposes.
First, it ascertains whether in this idealized setting,
the model parameters can be extracted uniquely
from short-term observations of the population dynamics.
Second, the analysis indicates how much information is required to estimate each model parameter accurately,
which yields valuable insights into how comparatively difficult it is to estimate the parameters
from more limited data.

\subsubsection{Cell number data} \label{sec:identifiabilitynum}

In the following proposition,
we show that for cell number data, 
the switching rates 
$(\nu_{ik})_{k \neq i}$
and the net birth rates $\boldsymbol\lambda$ can be recovered uniquely from knowledge of the mean functions ${\bf m}^{(j)}(t)$ close to time 0,
while the birth rates ${\bf b}$ can be recovered from the covariance matrices 
${\boldsymbol\Sigma}^{(j)}(t)$.

\begin{proposition} \phantomsection\label{cor:idcellnum}
\begin{enumerate}[(1)]
    \item For each $j=1,\ldots,K$, the switching rates $\nu_{jk}$, $k \neq j$, and the net birth rate $\lambda_j$ 
    are uniquely determined by 
    $\textstyle \frac{d}{dt} {\bf m}^{(j)}(t) \big|_{t=0}$.
    \item 
    For each $j=1,\ldots,K$, if the switching rates $\nu_{jk}$, $k \neq j$, and the net birth rate $\lambda_j$ are known,
    the birth rate $b_j$ is
    uniquely determined by 
    $\big(\textstyle \frac{d}{dt} {\boldsymbol\Sigma}^{(j)}(t) \big|_{t=0}\big)_{jj}$.
\end{enumerate}
\end{proposition}

\begin{proof}
Appendix \ref{app:identcellnum}.
\end{proof}

Proposition \ref{cor:idcellnum} establishes the structural identifiability of all model parameters for cell number data.
The process of extracting the parameters
as suggested by 
Proposition \ref{cor:idcellnum} 
can be 
thought of
as follows:
If we want to know $\nu_{jk}$ for some $k \neq j$, we can simply plot the mean function $M_{jk}(t) = \mathbb{E}\big[Z_k^{(j)}(t)\big]$
and compute its slope at 0.
If we want to know the birth rate $b_j$, we can plot the variance function $\textstyle \big({\boldsymbol\Sigma}^{(j)}(t) \big)_{jj} = {\rm Var}\big[Z_j^{(j)}(t)\big]$ and 
compute its slope at 0.

It is important to note that we are not suggesting to use this approach to estimate parameters from real data.
Instead, we are establishing theoretically that there is sufficient information in the distribution of the data close to time 0 to determine all model parameters uniquely.
In particular, we can in theory predict the entire evolutionary trajectory of the population
from short-term observations
of the 
initial
population dynamics.

\begin{table}
    \centering
\begin{tabular}{|c|c|c|c|}
\hline
    Moment & Derivative & Cell number data & Cell fraction data \\
    \hline 
    \multirow{2}{*}{1} & 1 & \,${\boldsymbol\lambda}$,\,$(\nu_{ik})_{k \neq i}$ & ${(\nu_{ik})_{k \neq i}}$ \\
     & 2 &  -- & ${\boldsymbol\lambda}^{[-1]}$ \\
        \hline
    2
    &1 & ${\bf b}$ &  $(\nu_{ik})_{k \neq i}$ \\
    \hline
\end{tabular}
\caption[Summary of the structural identifiability analysis of Propositions \ref{cor:idcellnum} and \ref{cor:idcellfrac1}.]{
Summary of the structural identifiability analysis of Propositions \ref{cor:idcellnum} and \ref{cor:idcellfrac1}.
For cell number data, the switching rates $(\nu_{ik})_{k \neq i}$ and the net birth rates $\boldsymbol\lambda$ are identifiable from the slopes (first derivatives) of the mean functions ${\bf m}^{(j)}(t)$ (first moments) at time 0.
The birth rates ${\bf b}$ are identifiable from the slopes of the covariance functions ${\boldsymbol\Sigma}^{(j)}(t)$ (second moments).
For cell fraction data, only the switching rates $(\nu_{ik})_{k \neq i}$ are identifiable from the slopes of the mean functions ${\bf p}^{(j)}(t)$, while the net birth rate differences $\boldsymbol\lambda^{[-1]}$ can be determined from their curvatures (second derivatives).
In contrast to cell number data, the slopes of the covariance functions ${\bf S}^{(j)}(t)$ for cell fraction data provide no extra information on the model parameters.
}
\label{table:identifiability}
\end{table}

\subsubsection{Cell fraction data} \label{sec:identifiabilitycellfraction}

In the following proposition, we show that for cell fraction data, only the switching rates $(\nu_{ik})_{k \neq i}$ can be recovered from the slopes of the mean functions ${\bf p}^{(j)}(t)$ at time 0.
The net birth rate differences $\boldsymbol\lambda^{[-1]}$ can be recovered from the curvatures of the mean functions at time 0 or from the equilibrium proportions $\overline{\boldsymbol\gamma}$ between cell types if they are known.
We are not able to learn any more parameters from the mean functions,
since ${\bf p}^{(j)}(t)$ can be written solely as a function of $(\nu_{ik})_{k \neq i}$ and $\boldsymbol\lambda^{[-1]}$.
The slopes of the covariance functions ${\bf S}^{(j)}(t)$ depend only on $(\nu_{ik})_{k \neq i}$, meaning that they provide no extra information on the model parameters.

\begin{proposition} \phantomsection\label{cor:idcellfrac1}
\begin{enumerate}[(1)]
    \item For $j=1,\ldots,K$, the switching rates $\nu_{jk}$, $k \neq j$, are uniquely determined by $\textstyle \frac{d}{dt} {\bf p}^{(j)}(t) \big|_{t=0}$.
    \item If the switching rates $(\nu_{ik})_{k \neq i}$ are known, the net birth rate differences $\boldsymbol\lambda^{[-1]}$ are uniquely determined by (i) $\textstyle \frac{d^2}{dt^2} {\bf p}^{(j)}(t) \big|_{t = 0}$ for $j=1,\ldots,K$ or (ii) the equilibrium proportions $\overline{\boldsymbol\gamma}$.
    \item For $j=1,\ldots,K$, $\textstyle \frac{d}{dt} {\bf S}^{(j)}(t) \big|_{t = 0}$ only depends on the switching rates $\nu_{jk}$ for $k \neq j$.
\end{enumerate}
\end{proposition}

\begin{proof}
Appendix \ref{app:fracident1}.
\end{proof}

As for the remaining model parameters, the net birth rate $\lambda_1$
and the birth rates ${\bf b}$,
they require information on the curvatures
of the covariance functions ${\bf S}^{(j)}(t)$ at time 0
at the least.
We will not analyze the structural identifiability of these parameters further.
Proposition \ref{cor:idcellfrac1} indicates that
one should not expect to be able to estimate these parameters accurately from cell fraction data,
which is confirmed by numerical experiments in Section \ref{sec:numericallargescale}.

\subsubsection{Comparison}

The results of our identifiability analysis are summarized in Table \ref{table:identifiability}.
Our analysis indicates that the switching rates $(\nu_{ik})_{k \neq i}$ and net birth rates $\boldsymbol\lambda$ are easy to estimate for cell number data, using information only on the mean behavior of the population.
The birth rates ${\bf b}$ are harder to estimate, since they require second moment information, but they may still be obtainable with sufficient data,
as we discuss further in Section \ref{sec:birthrates}.
For cell fraction data, the switching rates $(\nu_{ik})_{k \neq i}$ are easy to estimate using the mean behavior of the population.
The net birth rate differences $\boldsymbol\lambda^{[-1]}$ can also be estimated from the mean, but they require more information.
The remaining model parameters are unlikely to be obtainable from real datasets.

\subsection{Numerical experiments} \label{sec:numerical}

Next, we apply our maximum likelihood framework to computer-generated data.
In all cases, we assume that experiments are conducted from isolated initial conditions,
and we assume no measurement noise, i.e.~${\bf E}_{j,\ell}^{\rm num} = {\bf 0}$ and ${\bf E}_{j,\ell}^{\rm frac} = {\bf 0}$.
For simplicity, we only consider a model with two cell types, $K=2$.
Our goal is to assess how comparatively difficult it is to estimate the different model parameters depending on what data is collected.

\subsubsection{Implementation in MATLAB}

Our estimation framework has been implemented in MATLAB codes which are available at 
\url{https://github.com/egunnars/phenotypic_switching_inference/}.
The framework returns (i) a maximum likelihood estimate
and (ii) a likelihood-based confidence interval for each parameter,
using the sequential quadratic programming (sqp) solver in MATLAB.
Before solving the maximum likelihood problem,
we compute initial parameter estimates from a simpler model,
which we use to initialize the optimization
and to rescale the model parameters
so that they are of similar magnitude.
In most cases, we have found it sufficient to solve the maximum likelihood problem once,
starting from the simple estimates.
However, our MATLAB codes provide the option to solve the problem several times using different initial guesses.
Details of the implementation are provided in Appendix \ref{app:implementation}.

        \begin{figure}[!t]
        \centering
        \includegraphics[scale=1]{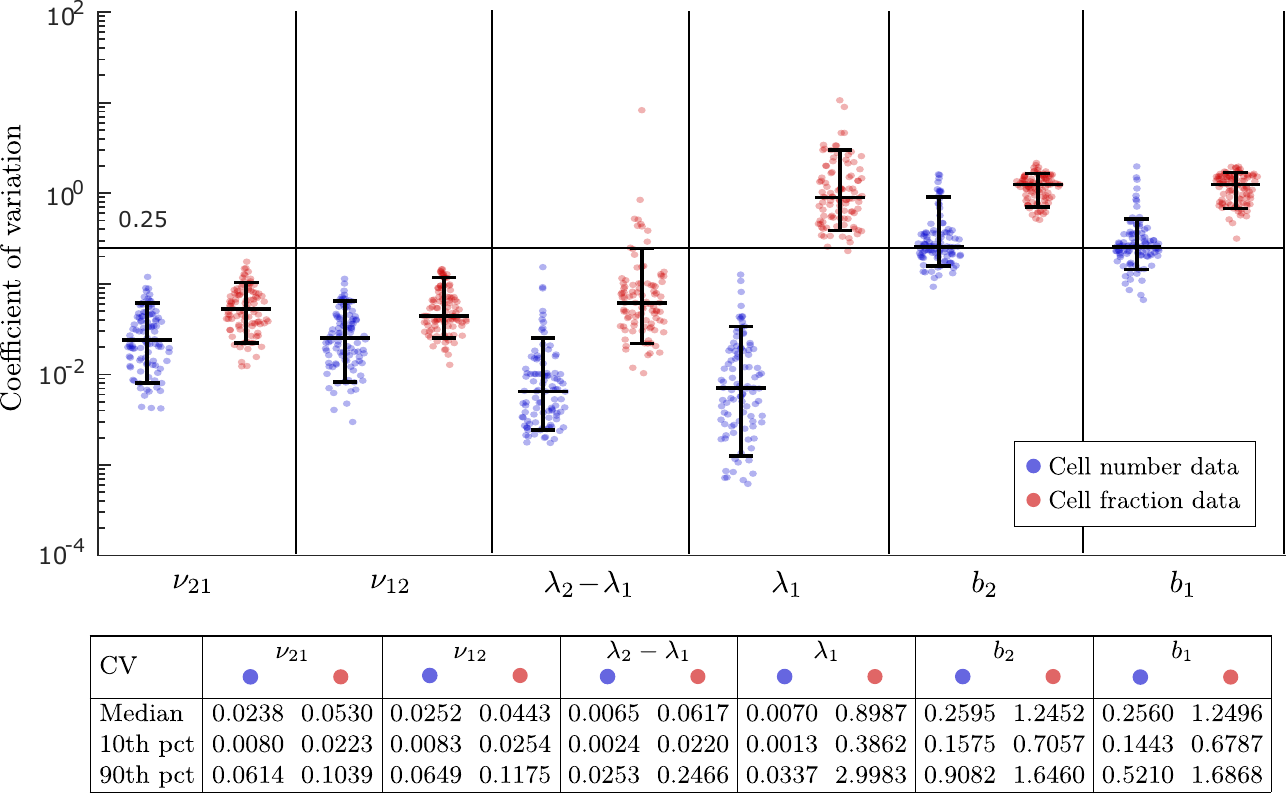}
        \caption[Assessment of estimation error
        across a wide range of biologically realistic parameter regimes.]{Assessment of estimation error
        across a wide range of biologically realistic parameter regimes.
        We first generated 100 different parameter regimes,
        then generated 100 artificial datasets for each regime,
        and finally computed parameter estimates for each dataset.
        To generate the parameter regimes, we sampled birth and death rates
        uniformly between 0 and 1, and sampled switching rates
        log-uniformly between $10^{-1}$ and $10^{-3}$ (Appendix \ref{app:numericalexp}).
        For each parameter and each parameter regime,
        we used the 100 estimates to compute the coefficient of variation (CV) for the estimates, which measures the error in the estimation.
        Each dot in the figure represents the CV for a single parameter under a single regime,
        with the blue dots (resp.~red dots) representing estimates from cell number data (resp.~cell fraction data).
        Collectively, the dots enable comparison of estimation error between different model parameters and between cell number and cell fraction data.
        The horizontal bars represent the 10th percentile, median and 90th percentile of the CVs, bottom to top. 
        }
        \label{fig:largescale}
    \end{figure}

\subsubsection{Estimation across a wide range of biologically realistic regimes} \label{sec:numericallargescale}

In Appendix \ref{sec:partexample}, we provide a simple illustration of 
the output of our estimation framework for a single artifical dataset.
For a more thorough evaluation of estimation accuracy,
we generated 10,000 artificial datasets for $K=2$ cell types.
We first generated 100 biologically realistic parameter regimes
and then generated 100 datasets for each regime.
To generate the parameter regimes, we sampled birth and death rates
uniformly between 0 and 1, and sampled switching rates
log-uniformly between $10^{-1}$ and $10^{-3}$.
We considered both regimes where the two phenotypes have positive net birth rates $(\lambda_1,\lambda_2 > 0)$ and regimes where one phenotype has a negative net birth rate $(\lambda_1 < 0,\lambda_2 > 0)$.
The latter regimes are relevant to the dynamics of anti-cancer treatment response, where one phenotype is drug-sensitive and the other is drug-tolerant or drug-resistant.
We assumed isolated initial conditions, $L=6$ time points and $R=3$ replicates.
Further details of the data generation are provided in Appendix \ref{app:numericalexp}.

For each dataset, we used our framework to compute MLE estimates for all model parameters.
In this way, we obtained 100 estimates of each parameter under each parameter regime, which we used to compute the coefficient of variation (CV) for the MLE estimator of the parameter.
The CV is the sample standard deviation of the MLE estimator as a proportion of its sample mean,
and it measures the percentage error in the estimation.

The results are shown in Figure \ref{fig:largescale}.
A horizontal line is drawn at 25\% CV to indicate whether
parameters can be estimated 
with reasonable accuracy.
Note that for the switching rates $(\nu_{ik})_{k \neq i}$, the median CV for cell fraction data is about twice as large as for cell number data.
The median CV for the net birth rate difference $\lambda_2-\lambda_1$ is an order of magnitude larger for cell fraction data than cell number data,
and it is two orders of magnitude larger for the net birth rate $\lambda_1$.
The birth rates ${\bf b}$ can in many cases be estimated reasonably well for cell number data,
whereas they are never estimated accurately for cell fraction data.
These results are very much in line with our identifiability analysis in Section \ref{sec:identifiability}.

Note that for cell fraction data,
the estimation error for the net birth rate difference $\lambda_2-\lambda_1$
exceeds the 25\% threshold CV for several parameter regimes.
This occurs when $\lambda_2-\lambda_1$ is small in magnitude,
more precisely when it is smaller than 0.1 in regimes
where the birth rates lie between 0.1 and 1.
Note in contrast that for cell number data,
the estimation error for $\lambda_2-\lambda_1$ never exceeds the 25\% threshold.
This indicates that for cell fraction data, 
it may be difficult to distinguish 
the net birth rate difference $\lambda_2-\lambda_1$ from 0
unless it is relatively pronounced.
We discuss this point further in Section \ref{sec:realdata} below.

In Appendix \ref{sec:expdesign},
we show how our framework can be 
used to investigate questions related to experimental design.
In particular, we consider the question
of whether experimental efforts
should be prioritized to collect data from more time points
(either in between or after the previous time points)
or to perform more experimental replicates.

        \begin{figure}[!t]
    \centering
    \includegraphics[scale=1]{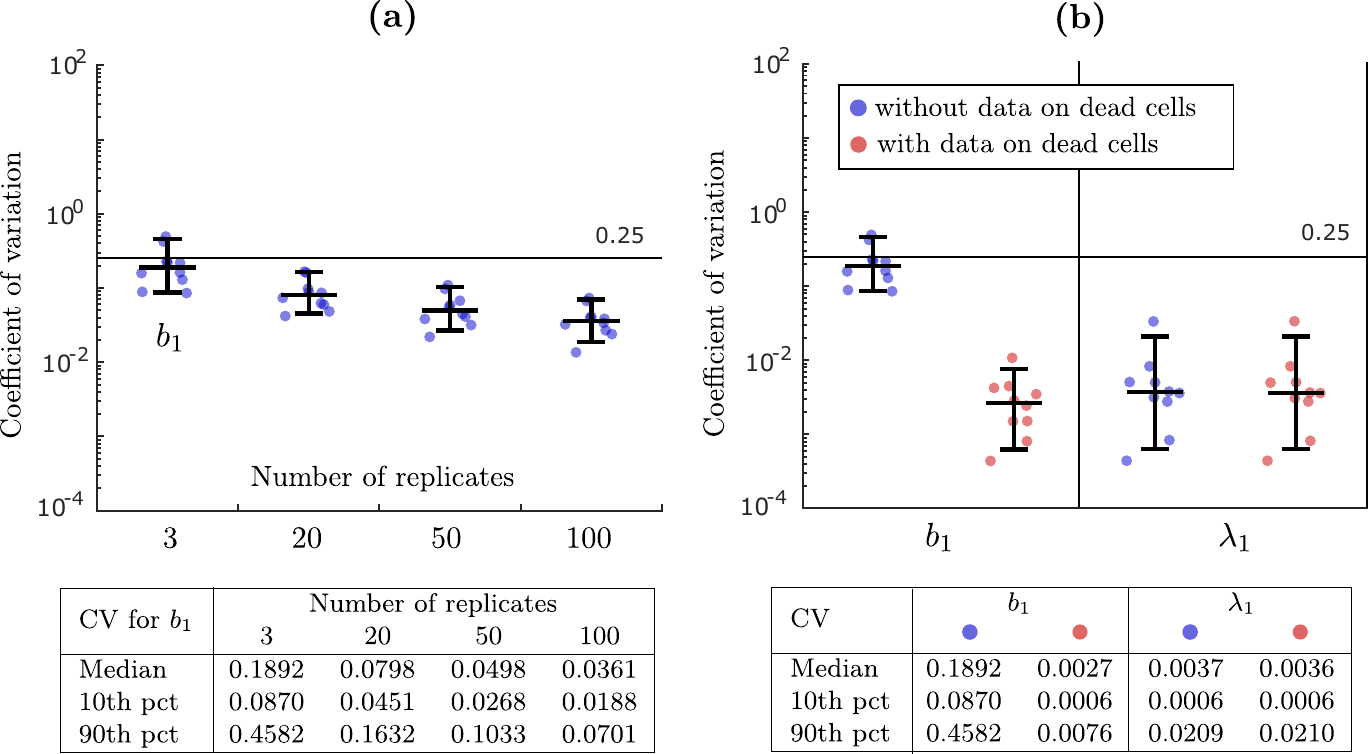}
    \caption[Two ways of improving the estimation accuracy for the birth rates ${\bf b}$ when cell number data is used.]{Two ways of improving the estimation accuracy for the birth rates ${\bf b}$ when cell number data is used.
    In {\bf (a)}, we show how the estimation accuracy for the birth rate $b_1$ improves as the number of experimental replicates is increased.
    In {\bf (b)}, we compare the estimation accuracy for the birth rate $b_1$ and the net birth rate $\lambda_1$ depending on whether data on the number of dead cells at each time point is included in the estimation or not.
    }
    \label{fig:birthrates}
\end{figure}

    \begin{figure}[!t]
    \centering
    \includegraphics[scale=0.8]{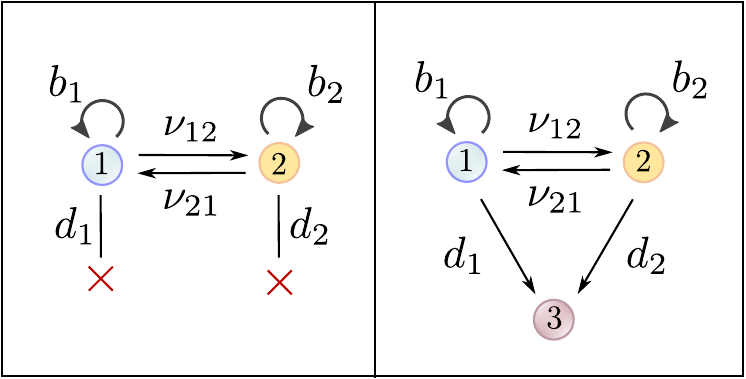}
    \caption[Augmentation of the mathematical model for when data is available on the number of dead cells at each time point.]{Augmentation of the mathematical model for when data is available on the number of dead cells at each time point.
    In that case, in stead of cells being lost from the model upon dying (left panel), they transition into a new state (right panel).
    }
    \label{fig:deadcellmodel}
\end{figure}

\subsubsection{Improving identifiability of the rates of cell division and cell death} \label{sec:birthrates}

For cell number data, even though the birth rates ${\bf b}$ can be estimated reasonably well in many cases by Section \ref{sec:numericallargescale}, they are estimated much less accurately than the net birth rates $\boldsymbol\lambda$ and the switching rates $(\nu_{ik})_{k \neq i}$.
In Figure \ref{fig:birthrates}a, we show that as the number of replicates is increased from 3 to 20 or above,
the accuracy in the estimation 
becomes more acceptable.
However, even with 100 replicates, the birth rates ${\bf b}$ are estimated less accurately
than the net birth rates $\boldsymbol\lambda$ with 3 replicates (see Figure \ref{fig:largescale}).

As we mentioned in the introduction,
data on the number of cells in each state at each time point
can 
be obtained by measuring the fraction of cells in each state
and the total number of cells at each time point.
In addition, it is often possible to measure the number of dead cells at each time point,
see e.g.~\cite{devaraj2019morphological}.
If 
this data is 
obtained,
we can augment our mathematical model by introducing a new cell state,
which cells transition into upon death (Figure \ref{fig:deadcellmodel}).
In Figure \ref{fig:birthrates}b, we show that if we apply our estimation framework to this model,
the birth rates ${\bf b}$ become as easy to estimate as the net birth rates $\boldsymbol\lambda$.
Thus, if data is collected 
on the number of live and dead cells at each time point,
it becomes possible to estimate all model parameters accurately
using our framework.

It should be noted that data collection on the number of dead cells is confounded by the fact that dead cells are eventually cleared from the system.
This can potentially be addressed by introducing a clearance rate
for dead cells in the augmented model,
i.e.~by introducing a death rate for the type-3 cells in the right panel of Figure \ref{fig:deadcellmodel}.

\begin{figure}[!t]
    \centering
    \includegraphics[scale=1]{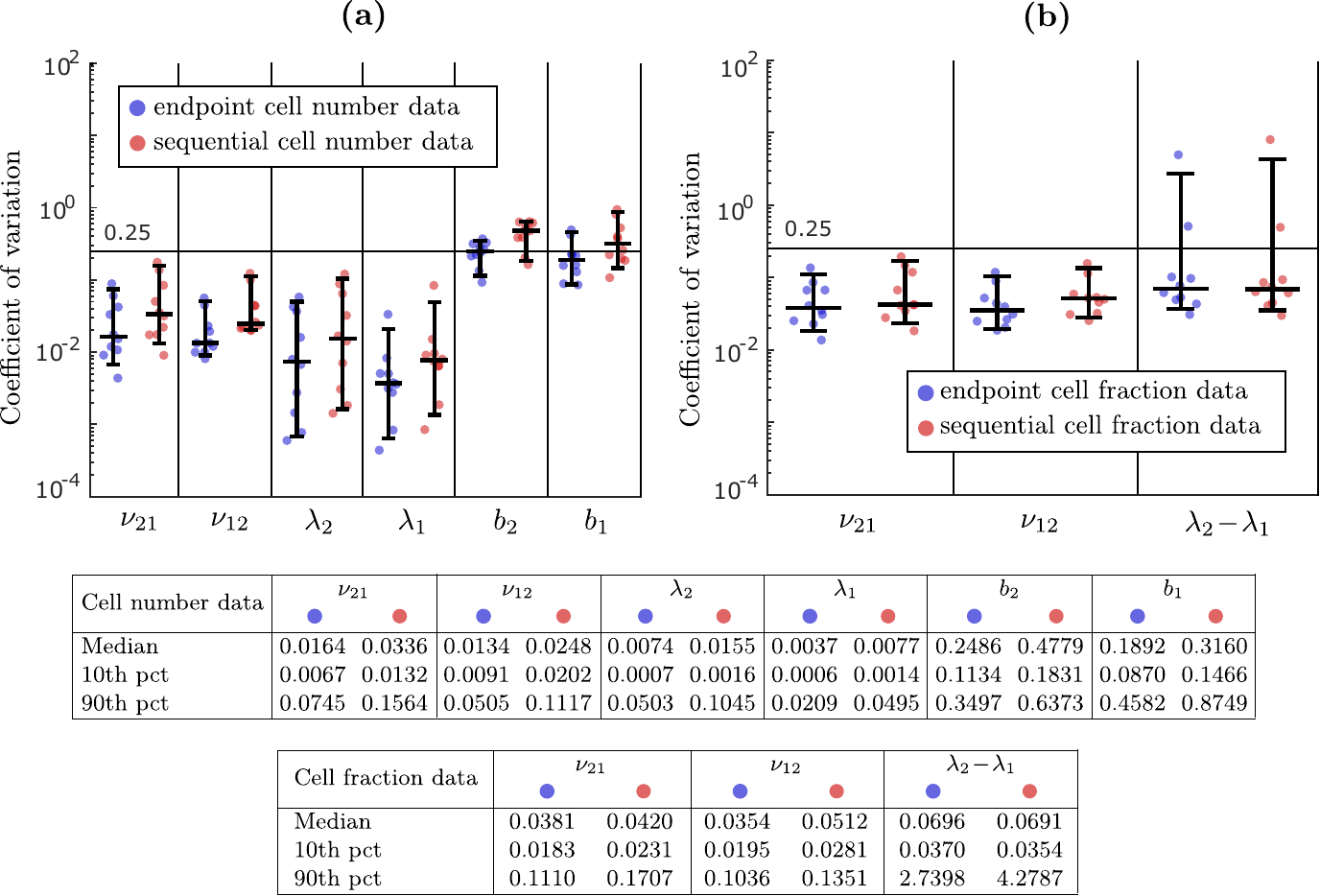}
    \caption[Comparison of estimation error depending on whether our framework is applied to endpoint data or sequential data.]{Comparison of estimation error depending on whether our framework is applied to endpoint data or sequential data.
    The blue dots show the estimation error when endpoint data is used,
    i.e.~when experiments from different time points are independent,
    and the red dots show the error when sequential data is used,
    i.e.~when data is collected at multiple time points in the same experiment.
    Panel {\bf (a)} shows the comparison for cell number data and panel {\bf (b)} for cell fraction data.
    Even though our framework is derived for endpoint data,
    it provides reasonable estimation accuracy for sequential data.
    }
    \label{fig:timelapse}
\end{figure}

\subsubsection{Estimation using endpoint data vs.~sequential data} \label{sec:endpointvssequential}

We conclude by examining how well our estimation framework
applies to 
sequential data, when data is collected at multiple time points in the same experiment (Section \ref{sec:experimentsanddata}).
In Figure \ref{fig:timelapse}, we see that for cell number data,
the CV for each parameter approximately doubles when 
applying our framework to sequential data vs.~endpoint data.
However, it remains true that the switching rates $(\nu_{ik})_{k \neq i}$
and net birth rates $\boldsymbol\lambda$ can be estimated with good accuracy.
For cell fraction data, the difference in the estimation error for $(\nu_{ik})_{k \neq i}$ and $\lambda_2-\lambda_1$ is even smaller.
Together, these results indicate that our framework
can yield reasonable estimates for sequential data.
At the same time, for cell number data in particular,
there can be a significant benefit to
developing a method tailored to sequential data,
both in terms of deriving reliable point estimates and robust confidence intervals.

\subsection{Application: Transition between stem and non-stem cell states in SW620 colon cancer} \label{sec:realdata}

To give an example of how our estimation
framework can be used to analyze real experimental data,
we conclude by applying it to
a publicly available cell fraction dataset.
We use data
collected by Yang et al.~\cite{yang2012dynamic} 
and made available in Tables S2 and S3 of Wang et al.~\cite{wang2014dynamics},
on the dynamics between stem-like (type-1) and non-stem (type-2) cells in SW620 
colon cancer.
In  
Yang et al.~\cite{yang2012dynamic},
the two cell types were sorted based on
expression of the CD133 cell-surface antigen marker.
Isolated subpopulations were expanded 
and 
phenotypic proportions were tracked for 24 days,
with data collected every other day.
This dataset has previously been analyzed using the CellTrans estimation method \cite{buder2017celltrans} (Section \ref{app:litreview}).

\begin{table}[t]
    \centering
\begin{tabular}{|c|c|c|c|c|c|c|}
\hline
    Model & ${\rm AIC}$ & ${\rm BIC}$ & $\nu_{21}$ & $\nu_{12}$ & $\lambda_2\!-\!\lambda_1$ \\
    \hline 
    \multirow{2}{*}{I}& 
    \multirow{2}{*}{$-113.4$}&\multirow{2}{*}{$-105.2$}
    & 0.154 & 0.057 & 0.080  \\
    & & &  CI: $[0.111,0.212]$ & CI: $[0.036,0.087]$ & CI: $[-0.058,0.219]$ \\
    \hline
     \multirow{2}{*}{Ia}& 
      \multirow{2}{*}{$-119.4$}&\multirow{2}{*}{$-114.7$}
      & 0.157 & 0.057 & 0.084 \\
          & & &  CI: $[0.115,0.213]$ & CI: $[0.037,0.088]$ & CI: $[-0.054,0.218]$ \\
    \hline
\end{tabular}
\caption{
Comparison of model fit quality, point estimates and confidence intervals for the statistical models ${\bf f}_{j,\ell} \;{\sim}\; {\bf p}^{(j)}(t_\ell) + {\cal N}\big({\bf 0},N_j^{-1}{\bf S}^{(j)}(t_\ell)\big) + {\cal N}\big({\bf 0}, \omega^2{\bf I}\big)$ (Model I) and ${\bf f}_{j,\ell} \;{\sim}\; {\bf p}^{(j)}(t_\ell) + {\cal N}\big({\bf 0}, \omega^2{\bf I}\big)$ (Model Ia) applied to publicly available cell fraction data from Yang et al.~\cite{yang2012dynamic}.
}
\label{table:realdata1}
\end{table}

Since data on individual experimental replicates is not available,
we use data on the mean cell fraction across replicates 
as input to our estimation framework.
We first consider the statistical model
\eqref{eq:statisticalmodelfrac} and the simpler version \eqref{eq:statisticalmodelfracsimple}
with ${\bf E}^{\rm frac}_{j,\ell} = \omega^2{\bf I}$ for all $j,\ell$, which we refer to as Models I and Ia, respectively:
\begin{itemize}
    \item {\bf Model I:} ${\bf f}_{j,\ell} \;{\sim}\; {\bf p}^{(j)}(t_\ell) + {\cal N}\big({\bf 0},N_j^{-1}{\bf S}^{(j)}(t_\ell)\big) + {\cal N}\big({\bf 0}, \omega^2{\bf I}\big)$.
    \item {\bf Model Ia:} ${\bf f}_{j,\ell} \;{\sim}\; {\bf p}^{(j)}(t_\ell) + {\cal N}\big({\bf 0}, \omega^2{\bf I}\big)$.
\end{itemize}
Note that Model I has seven parameters $(d_1,d_2,\lambda_1,\lambda_2-\lambda_1,\nu_{12},\nu_{21},\omega)$, while Model Ia has four parameters ($\lambda_2-\lambda_1,\nu_{12},\nu_{21},\omega)$.
In Table \ref{table:realdata1}, we show parameter estimates and 95\% confidence intervals for the two models, which turn out to be very similar.
By the Akaike Information Criterion (AIC) and the Bayesian Information Criterion (BIC), 
which assess the quality of model fit relative to model complexity, 
the simpler Model Ia is preferred for this dataset (Appendix \ref{app:AIC}).
The codes used to compute the estimates in Table \ref{table:realdata1} are available at \url{https://github.com/egunnars/phenotypic_switching_inference/}.

The CIs under Model Ia show that while the point estimates for $\nu_{21}$ and $\nu_{12}$ are 0.157 and 0.057, respectively,
the true value of $\nu_{21}$ may range between 0.115 and 0.213,
and the true value of $\nu_{12}$ may range between 0.037 and 0.088.
Since the two CIs do not overlap,
$\nu_{21} > \nu_{12}$ at the 5\% level of significance,
but there is considerable uncertainty as to the true values.
The CI for $\lambda_2-\lambda_1$ is even wider,
which 
is in line with our earlier observations that this parameter is more difficult to estimate from cell fraction data than the switching rates,
especially when $\lambda_2-\lambda_1$ is relatively small in magnitude (Sections \ref{sec:identifiabilitycellfraction} and \ref{sec:numericallargescale}).
In fact, the CI for $\lambda_2-\lambda_1$ 
includes zero,
meaning that it is plausible that $\lambda_1=\lambda_2$.

\begin{figure}[!t]
    \centering
    \hspace*{-27pt} \includegraphics[scale=0.8]{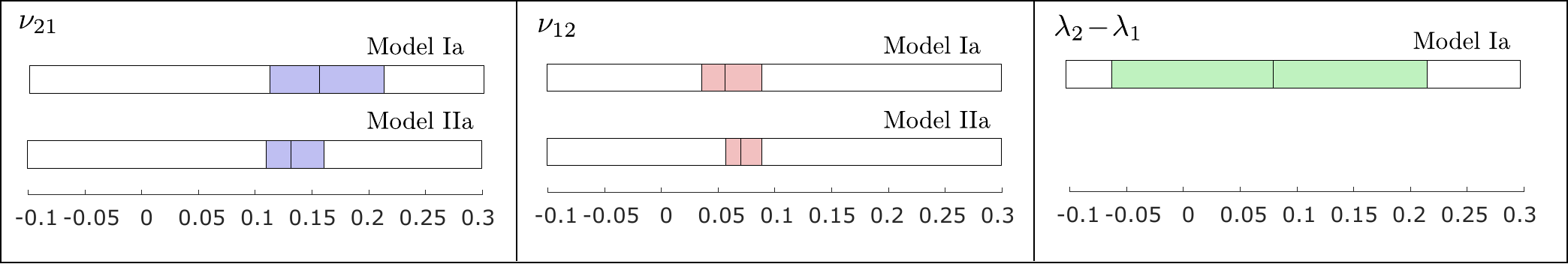}
\caption{
Visual comparison of point estimates and 95\% confidence intervals for the statistical model ${\bf f}_{j,\ell} \;{\sim}\; {\bf p}^{(j)}(t_\ell) + {\cal N}\big({\bf 0}, \omega^2{\bf I}\big)$ (Model Ia) and the same model with $\lambda_2-\lambda_1=0$ (Model IIa) applied to publicly available cell fraction data from Yang et al.~\cite{yang2012dynamic}.}
    \label{fig:realdata2}
\end{figure}

In the CellTrans paper \cite{buder2017celltrans}, it is assumed that the two phenotypes have the same growth rate, based on data from Wang et al.~\cite{wang2014dynamics}.
We can build this assumption
into the estimation by 
solving the MLE problem for Models I/Ia
under
the constraint $\lambda_2-\lambda_1=0$ (Appendix \ref{app:implementation}).
We refer to this as Models II/IIa:
\begin{itemize}
        \item {\bf Model II}: ${\bf f}_{j,\ell} \; {\sim} \; {\bf p}^{(j)}(t_\ell) + {\cal N}\big({\bf 0},N_j^{-1}{\bf S}^{(j)}(t_\ell)\big) + {\cal N}\big({\bf 0}, \omega^2{\bf I}\big)$, $\lambda_2-\lambda_1=0$.
        \item {\bf Model IIa}: ${\bf f}_{j,\ell} \; {\sim} \; {\bf p}^{(j)}(t_\ell) + {\cal N}\big({\bf 0}, \omega^2{\bf I}\big)$, $\lambda_2-\lambda_1=0$.
\end{itemize}
Estimation results for Models II/IIa are shown in Table \ref{table:realdata2},
and a visual comparison between the estimates for Models Ia and IIa is shown in Figure \ref{fig:realdata2}.
The assumption $\lambda_1=\lambda_2$ has a noticeable effect on both the point estimates of $\nu_{21}$ and $\nu_{12}$ and their confidence intervals.
For example, the ratio $\nu_{21}/\nu_{12}$ 
is 2.7 under Model Ia, while it is 1.9 under Model IIa.
In other words, switching from type-2 to type-1 happens about three times as often as switching from type-1 to type-2 under Model Ia, while it happens about two times as often under Model IIa.
Furthermore, under Model IIa, the length of the CI for $\nu_{21}$ is reduced by a half compared to Model Ia,
meaning that Model IIa significantly restricts the plausible values of $\nu_{21}$.

\begin{table}[t]
    \centering
\begin{tabular}{|c|c|c|c|c|c|}
\hline
    Model & ${\rm AIC}$ & ${\rm BIC}$ & $\nu_{21}$ & $\nu_{12}$ \\
    \hline 
    \multirow{2}{*}{II}& 
    \multirow{2}{*}{$-114.0$}&\multirow{2}{*}{$-107.0$}
    & 0.131 & 0.071   \\
    & & &  CI: $[0.110,0.161]$ & CI: $[0.057,0.089]$   \\
    \hline
     \multirow{2}{*}{IIa}& 
      \multirow{2}{*}{$-119.8$}&\multirow{2}{*}{$-116.3$}
      & 0.134 & 0.072  \\
          & & &  CI: $[0.112,0.162]$ & CI: $[0.059,0.090]$  \\
    \hline
\end{tabular}
\caption{
Comparison of model fit quality, point estimates and confidence intervals for the statistical models ${\bf f}_{j,\ell} \;{\sim}\; {\bf p}^{(j)}(t_\ell) + {\cal N}\big({\bf 0},N_j^{-1}{\bf S}^{(j)}(t_\ell)\big) + {\cal N}\big({\bf 0}, \omega^2{\bf I}\big)$, $\lambda_2-\lambda_1=0$ (Model II) and ${\bf f}_{j,\ell} \;{\sim}\; {\bf p}^{(j)}(t_\ell) + {\cal N}\big({\bf 0}, \omega^2{\bf I}\big)$, $\lambda_2-\lambda_1=0$ (Model IIa) applied to publicly available cell fraction data from Yang et al.~\cite{yang2012dynamic}.
}
\label{table:realdata2}
\end{table}

In the CellTrans paper \cite{buder2017celltrans}, the same dataset is used to estimate
switching probabilities of $p_{21} = 0.1030$ and $p_{12} = 0.0545$,
based on a discrete-time Markov model with a time step of $\Delta t =$ one day.
We also solved the TRANSCOMPP problem \eqref{eq:TRANSCOMPP} (see Section \ref{app:litreview}) with $\Delta t =$ one day to obtain the estimates ${p}_{21} = 0.136$ and ${p}_{12} = 0.054$ for the switching probabilities and $\Lambda_{22}/\Lambda_{11} = 1.079$ for the ratio between the growth factors of the two phenotypes,
which
translates to a growth rate difference of $r_2-r_1 = 0.076$ if we set $\Lambda_{22} = e^{r_2 \Delta t}$ and $\Lambda_{11} = e^{r_1 \Delta t}$.
In the CellTrans and TRANSCOMPP models, type switches are synchronized between all cells in the population, and they occur at discrete time steps.
In our continuous-time model,
the time steps are infinitesimally small,
and each cell has a certain probability of switching, proliferating
and dying during each step, independently of other cells
(Section \ref{sec:modelparameters}).
If we shorten the time step 
to $\Delta t = 1/10$ day,
the switching probabilities become $0.0111$ and $0.0059$ under CellTrans,
which translates to continuous-time rates of $\widetilde{p}_{21} = 0.111$ and $\widetilde{p}_{12} = 0.059$.
These estimates fall at the lower limits of our CIs for $\nu_{21}$ and $\nu_{12}$ under Models II/IIa (Table \ref{table:realdata2}).
Under TRANSCOMPP,
the switching probabilities become 0.0154 and 0.0057 for $\Delta t = 1/10$ day,
which translates to continuous-time rates of $\widetilde{p}_{21} = 0.154$ and $\widetilde{p}_{12} = 0.057$, and the difference in growth rates becomes $\widetilde r_2-\widetilde r_1 = 0.083$.
These estimates are very similar to the point estimates of Models I/Ia (Table \ref{table:realdata1}).

The estimates of CellTrans and TRANSCOMPP are consistent with our estimates
in that they fall within the 95\% confidence intervals produced by our framework,
if the time step is taken to be sufficiently small.
Our framework complements these methods
for cell fraction data
by providing continuous-time estimates and
enabling a rigorous analysis
of the estimates and the uncertainty involved.
For example, the CIs provided by our framework reveal how
uncertain the value of $\lambda_2-\lambda_1$ is
compared to $\nu_{21}$ and $\nu_{12}$,
and that $\lambda_2-\lambda_1$ cannot be distinguished from zero
using this dataset.
If assumptions such as
$\lambda_1 = \lambda_2$ or $\nu_{12} = \nu_{21}$
can be made,
it is easy to incorporate them
into the estimation
and to assess 
their effect
on point estimates and confidence intervals (Appendix \ref{app:implementation}).
In this case, our analysis shows that
the assumption $\lambda_1=\lambda_2$ significantly restricts the plausible
values of $\nu_{21}$ and $\nu_{12}$,
which may underestimate the true 
uncertainty in the estimation,
given that the claim $\lambda_1=\lambda_2$
is subject to statistical error.
We discuss the differences between our approach and these two methods,
and the importance of quantifying the uncertainty in the estimation,
in more detail in the following section.

\section{Discussion} \label{sec:discussion}

In this work, we have proposed a 
maximum likelihood framework for estimating the rates of cell proliferation and phenotypic switching in cancer.
In contrast to previous approaches, 
the framework
explicitly models the stochastic dynamics of cell division, cell death and phenotypic switching,
it provides likelihood-based confidence intervals for the model parameters,
and it enables estimation from data
on the fraction of cells or the number of cells
in each state at each time point.
An implementation of the framework in MATLAB with sample scripts is available at \url{https://github.com/egunnars/phenotypic_switching_inference/}.

We have also used our framework to analyze the identifiability of model parameters.
Through a combination of theoretical and numerical investigation
and application to real data,
we have seen that when cell fraction data is used,
the switching rates $(\nu_{ik})_{k \neq i}$
may be the only parameters that can be estimated accurately,
while the net birth rate differences $\boldsymbol\lambda^{[-1]}$
can also be estimated reasonably accurately if they are sufficiently large.
Including information on the total size of the population at each time point yields significantly better estimates of $\boldsymbol\lambda^{[-1]}$,
and it also enables accurate estimation of the net birth rates $\boldsymbol\lambda$.
Finally, if enough experimental replicates are performed,
or if data is collected on the number of dead cells at each time point,
it even becomes possible to estimates the birth rates ${\bf b}$ and death rates ${\bf d}$ accurately.

In a previous work, we discussed how knowledge of
the 
model 
parameters $(\nu_{ik})_{k \neq i}$, $\boldsymbol\lambda$, ${\bf b}$
can enhance our understanding of resistance evolution in cancer
and inform the design of combination treatments of anti-cancer agents
and epigenetic drugs \cite{gunnarsson2020understanding}.
Together, these parameters shape the evolution
of phenotypic proportions and the total tumor burden over time,
each of which is relevant to
the dynamics of tumor recurrence.
Our current work shows that it is not possible to estimate
the net birth rates $\boldsymbol\lambda$
or the birth rates ${\bf b}$
accurately from cell fraction data,
it indicates what data is required
to obtain these parameters,
and it offers a rigorous
approach to parameter 
estimation and uncertainty quantification
once the data has been acquired.
In the context of anti-cancer drug response,
uncertainty quantification is crucial for assessing
how treatment affects the model parameters
and for evaluating the robustness of any treatment recommendations.
For example, there is evidence that both chemotherapies and targeted agents
can induce phenotypic switching from drug-sensitivity to drug-tolerance \cite{goldman2015temporally,su2017single,russo2022modified},
where the level of induction 
determines the optimal dose
under continuous drug treatment
\cite{greene2019mathematical,kuosmanen2021drug,angelini2022model}.
In this context, it is important to confirm that an estimated induction of drug-tolerance is statistically significant, and to assess how accurately the induction level can be estimated,
before dose changes for established treatment protocols can be recommended.

In our application to a publicly available cell fraction dataset,
we compared estimates from our framework to estimates produced by
CellTrans \cite{buder2017celltrans} and TRANSCOMPP \cite{jagannathan2020transcompp}.
CellTrans is based on a discrete-time Markov chain model, and it provides estimates for the probabilities of switching between phenotypes during a single time step, for the case where all types grow at the same rate.
TRANSCOMPP is based on a similar model, except it also provides estimates of the relative growth rates of the different phenotypes,
and it includes a stochastic resampling method for estimating the distributions of transition probability estimates
using single-cell measurements.
Each method can only be applied to cell fraction data.
For the 
dataset studied in Section \ref{sec:realdata},
CellTrans and TRANSCOMPP produce estimates similar to our framework,
when the time step is taken to be sufficiently small.
We expect that this will usually be the case for datasets with few experimental replicates
or a large measurement error, since our baseline statistical model \eqref{eq:statisticalmodelfrac} incorporates second moment information which is likely to be distorted in such datasets.
However, we believe that even for these datasets,
the continuous-time estimates provided by our framework  
better reflect the asynchronous nature of cell state switching, division and death,
and they have the benefit of
not being affected 
by an arbitrary choice of time step.
More importantly, our framework provides likelihood-based confidence intervals for 
the parameters
$(\nu_{ik})_{k \neq i}$ and 
$\boldsymbol\lambda^{[-1]}$, which is crucial to assess the quality of the estimation.
Finally, our framework is unique in that it enables estimation from cell number data.
It should be noted that for cell number data in particular, the appropriate measurement error model 
may vary between specific applications, as is discussed below.

There are several limitations of the estimation framework,
which represent avenues for future development and improvement.
First, our framework assumes that the cell population can be decomposed
into discrete phenotypes, which can be identified 
using known biomarkers.
Second, our multitype branching process model assumes that
the lifetime of a cell is exponentially distributed,
meaning that the rate at which a cell divides or dies
is independent of its age.
It is possible to model non-exponential lifetimes using our framework
by assuming that each phenotype transitions through
a number of internal states, each at an exponential rate, before dividing or dying.
This will however increase the number of parameters in the model,
which will require more data to obtain accurate estimates.
Another approach would be to employ age-dependent branching processes,
which would also add parameters to the model
\cite{athreya2004branching}.
A third limitation of our framework is that it ignores any potential cell-to-cell interactions.
Incorporating such interactions likely requires estimation methods tailored to specific applications, depending on the specific nature of the interactions.

Fourth, the branching process model assumes that
cells are allowed to grow uninterrupted for the duration of the experiments.
This does not address the effect of passaging
in longer-duration experiments.
One potential way to address passaging
is to keep track of cell state proportions
and seeding densities for each passage,
and to consider each passage as a new
experiment with new initial conditions.
In other words, instead of viewing a long experiment involving 
serial passaging
as a single experiment with a single initial condition,
it can be viewed as a collection of shorter experiments
with different initial conditions.
However, our framework currently assumes that initial conditions are 
known,
while uncertainty is assigned to all subsequent time points.
In reality, the initial conditions are subject to measurement error,
and it may become important to 
model this error for the case of repeated passaging.

Fifth, our framework currently models measurement error
as an additive Gaussian noise
with a general covariance matrix.
We have suggested simple ways of choosing the covariance matrix
both for cell number and cell fraction data,
but further exploration of appropriate choices is warranted.
Ideally, the determination of an appropriate measurement error model should
be driven by the particular dataset being analyzed \cite{benzekry2014classical}.
Depending on the application,
it may 
also 
become necessary to develop a more sophisticated
error model than the additive Gaussian model.
For example, for cell number data,
if the measurement error is proportional to the population size,
it may become necessary to model it as a multiplicative term rather than an additive term,
or to build the experimental
cell counting procedure  
more explicitly into the statistical model.
We plan to address this in future work.

Sixth, we have focused 
on estimation
from experiments started with isolated subpopulations
of each phenotype,
as this is a common experimental design,
and we have analyzed
parameter identifiability
in this setting.
Understanding to what extent the model parameters,
or some combinations of the parameters,
can be estimated from more limited data
is an interesting avenue for future investigation.
For example, if we are interested in estimating
parameters from clinical data,
the data will likely contain much
less information than we have assumed here,
and it will become necessary to analyze what parameters are identifiable
and how identifiability can be improved,
e.g.~by combining data from similar patients.

Finally, 
we believe our framework can be useful
for the design of cell line experiments 
aimed at deciphering the dynamics of phenotypic switching.
For example, preliminary experiments can first be conducted,
from which initial parameter estimates and confidence intervals
can be derived.
Based on the confidence intervals,
one can construct a set of likely values for the parameters,
which can be used to evaluate the expected 
improvement 
in estimation accuracy
depending on the experimental design
(see e.g.~\cite{steiert2012experimental}).
Once good experimental designs have been identified,
one can evaluate whether the expected improvement in estimation accuracy
justifies the additional experimental resources.
If this is the case,
additional experiments can be performed and the process can be repeated.
In a future work, we plan to develop a tool for the optimal selection of experimental designs,
to facilitate
more efficient utilization of 
experimental resources.

\newpage

\appendix

\section{Estimation framework} \label{app:estimationdeets}

In Sections \ref{sec:estimationnum} and \ref{sec:estimationfrac} of the main text,
we described our estimation framework for the simple case
where all experiments are started from isolated subpopulations.
We also omitted the details regarding the computation of 
point estimates and confidence intervals.
In this section, we develop the estimation framework in full detail
for general starting conditions.

\subsection{Notation for experimental data} \label{app:experimentsanddata}

For the general case, we assume that each experiment is started with a known initial condition,
encoded by the $1 \times K$ vector ${\bf n}=(n_1,\ldots,n_K)$ of starting cell numbers of each type.
We let $I \geq 1$ denote the number of distinct initial conditions
and
${\bf n}_i = (n_{i1},\ldots,n_{iK})$ denote the $i$-th initial condition.
We  assume that 
for each $i=1,\ldots,I$ and $j=1,\ldots,K$, either $n_{ij}=0$ or $n_{ij}$ is large, which is
generally the case for the experiments discussed in the introduction (Section \ref{sec:introduction}).

We define $N_i := \sum_{k=1}^K n_{ik}$ as the total number of starting cells in the $i$-th condition, and ${\bf f}_i = (f_{i1},\ldots,f_{iK})$ as the vector of starting cell fractions, 
with $f_{ij} := n_{ij}/N_{i}$.
As in Section \ref{sec:experimentsanddata} of the main text, we let $L \geq 1$
be the number of time points at which data is collected,
and we let $0 < t_1 < t_2 < \cdots < t_{L}$ denote the 
time points.
Finally, we let $R \geq 1$ be the number of experimental replicates performed.

The data collected in each experiment is either a vector ${\bf n}_{i,\ell,r} = (n_{i,\ell,r,1},\ldots,n_{i,\ell,r,K})$ of cell numbers or ${\bf f}_{i,\ell,r} = (f_{i,\ell,r,1},\ldots,f_{i,\ell,r,K})$ of cell fractions.
Here, $n_{i,\ell,r,k}$ is the number of type-$k$ cells in the $r$-th replicate of the experiment started by the $i$-th initial condition and ended at the $\ell$-th timepoint,
and $f_{i,\ell,r,k}$ is the corresponding cell fraction.

\subsection{Estimation for cell number data} \label{app:estimationcellnum}

We now develop the estimation framework for cell number data.
For the general case, the starting vector ${\bf f}_{i}$ of cell fractions can be any $1 \times K$ vector $\boldsymbol\alpha$ with $\alpha_k \geq 0$ for $k=1,\ldots,K$ and $\sum_{k=1}^K \alpha_k=1$.
In expression \eqref{eq:mSigma} of the main text, we defined the mean function ${\bf m}^{(j)}(t)$ and the covariance matrix ${\boldsymbol\Sigma}^{(j)}(t)$ for an isolated initial condition.
We extend these definitions to a general vector $\boldsymbol\alpha$ of starting cell fractions as follows:
\begin{align} \label{eq:malphaSigmaalpha}
\begin{split}
        & \textstyle {\bf m}^{\boldsymbol\alpha}(t) := {\boldsymbol\alpha}{\bf M}(t) = \sum_{j=1}^K \alpha_j {\bf m}^{(j)}(t), \\
    & \textstyle {\boldsymbol\Sigma}^{\boldsymbol\alpha}(t) := \sum_{j=1}^K \alpha_j {\boldsymbol\Sigma}^{(j)}(t).
\end{split}
\end{align}
Then, based on a generalized version of the central limit theorem \eqref{eq:CLTnum},
which is stated and proved as Proposition \ref{prop:CLTnum} in Appendix \ref{app:CLT1},
we propose the following extension of the statistical model \eqref{eq:statisticalmodelnum} in the main text:
\begin{align} \label{eq:statisticalmodelnumgeneral}
        {\bf n}_{i,\ell,r} 
    \; \sim \;
    \underbrace{N_i {\bf m}^{{\bf f}_i}(t_\ell)}_{\substack{{\rm mean} \\{\rm behavior}}} \;+\; \underbrace{{\cal N}\big({\bf 0}, N_i {\boldsymbol\Sigma}^{{\bf f}_i}(t_\ell)\big)}_{\substack{{\rm variability \; in} \\ {\rm population \; dynamics}}} \;+\; \underbrace{{\cal N}\big({\bf 0}, {\bf E}_{i,\ell}^{\rm num}\big)}_{\substack{{\rm measurement} \\{\rm error}}}.
\end{align}
The vectors ${\bf n}_{i,\ell,r}$ and ${\bf n}_{j,m,s}$ are assumed independent for
$(i,\ell,r) \neq (j,m,s)$, and they are assumed i.i.d.~for $(i,\ell) = (j,m)$ and $r \neq s$.
This implies
that data from distinct time points
come from distinct experiments (endpoint data).
We assume endpoint data since the central limit theorem (CLT) in Proposition \ref{prop:CLTnum} 
holds for the distribution of cell numbers at a fixed time point $t$.
Developing an analogous statistical model for sequential data
requires extending the CLT to a process-level or functional CLT.
We plan to address this in future work.

Note that in the statistical model \eqref{eq:statisticalmodelnumgeneral}, 
the mean behavior $N_i {\bf m}^{{\bf f}_i}(t_\ell)$ of the model
depends only on the switching rates $(\nu_{ik})_{k \neq i}$ and the net birth rates 
${\boldsymbol\lambda}$,
while the variance term $N_i {\boldsymbol\Sigma}^{{\bf f}_i}(t_\ell)$ 
depends on $(\nu_{ik})_{k \neq i}$, ${\boldsymbol\lambda}$ and also the birth rates ${\bf b}$
by \eqref{eq:covmatrixexpression}.
It is therefore natural to 
parametrize the 
first two terms in \eqref{eq:statisticalmodelnumgeneral} 
by
${\bf b},\boldsymbol\lambda,(\nu_{ik})_{k \neq i}$
instead of the primary model parameters
${\bf b},{\bf d},(\nu_{ik})_{k \neq i}$.
As stated in the main text, we assume that the $K \times K$ covariance matrix ${\bf E}_{i,\ell}^{\rm num}$ associated with measurement error can be written as a function 
of ${\bf b},\boldsymbol\lambda,(\nu_{ik})_{k \neq i}$
and added error parameters $\boldsymbol\omega_{\rm num} = (\omega_{1},\ldots,\omega_{M_{\rm num}})$
for some $M_{\rm num} \geq 0$.
We let ${\boldsymbol\theta}_{\rm num}$ be the complete $1 \times \big(K(K+1)+M_{\rm num}\big)$ 
vector of model parameters including the error parameters.

From the statistical model \eqref{eq:statisticalmodelnumgeneral},
it is straightforward to derive the following likelihood function:
    \begin{align} \label{eq:likelihoodnum}
    &{\cal L}_{\rm num}\big(
    {\boldsymbol\theta}_{\rm num}
    \big|
    (
    {\bf n}_{i,\ell,r})
    \big) \nonumber \\
    &= \textstyle \prod_{i=1}^I \prod_{\ell=1}^{L} \prod_{r=1}^{R} \! \Big((2\pi)^{K} {\rm det}\big( N_i{\boldsymbol\Sigma}^{{\bf f}_i}(t_\ell)+{\bf E}_{i,\ell}^{\rm num}\big)\Big)^{-1/2} \nonumber \\
    &\quad \textstyle \cdot\exp\big(\!-\!\frac12 \big({\bf n}_{i,\ell,r}-N_i{\bf m}^{{\bf f}_i}(t_\ell)\big) \big(N_i{\boldsymbol\Sigma}^{{\bf f}_i}(t_\ell)+{\bf E}_{i,\ell}^{\rm num}\big)^{-1} \big({\bf n}_{i,\ell,r}-N_i{\bf m}^{{\bf f}_i}(t_\ell)\big)^T\big).
\end{align}
We next define the 
negative double log-likelihood,
\begin{align} \label{eq:loglikelihood}
    & l_{\rm num}\big(
    {\boldsymbol\theta}_{\rm num}
    \big) 
    := -2\log {\cal L}_{\rm num} \big({\boldsymbol\theta}_{\rm num}
    \big|({\bf n}_{i,\ell,r})
    \big).
\end{align}
The maximum likelihood estimate $\widehat{\boldsymbol\theta}_{\rm num}$ for the 
parameter vector $\boldsymbol\theta_{\rm num}$
is obtained by minimizing 
$l_{\rm num}\big(
    {\boldsymbol\theta}_{\rm num}\big)$
over a set of feasible parameters $\boldsymbol\Theta_{\rm num}$:
\begin{align} \label{eq:mleestimator}
    \widehat{\boldsymbol\theta}_{\rm num}
    := {\rm argmin}_{\boldsymbol\theta_{\rm num} \in \boldsymbol\Theta_{\rm num}} \, 
    l_{\rm num}(\boldsymbol\theta_{\rm num}).
\end{align}
In the feasible set $\boldsymbol\Theta_{\rm num}$, we restrict the parameter values so that $\nu_{ik} \geq 0$,
${\bf b} \geq {\bf 0}$ and ${\boldsymbol\lambda} \leq {\bf b}$.
Further restrictions can be made depending on the context, see e.g.~Appendix \ref{sec:extensions}.

A $1-\alpha$ likelihood-based confidence interval $ \big[\theta_{{\rm num},i}^-,\theta_{{\rm num},i}^+\big]$ for the $i$-th model parameter $\theta_{{\rm num},i}$ can be obtained by collecting all values $\theta$ for which the null hypothesis $\theta_{{\rm num},i} = \theta$ is accepted under the likelihood-ratio test.
To describe how the confidence interval is obtained,
we 
define the
negative double {\em profile log-likelihood}
for $\theta_{{\rm num},i}$ as
\[
\widetilde{l}_{{\rm num},i}(\theta) := \min_{\boldsymbol\theta_{\rm num} \in \boldsymbol\Theta_{\rm num}: \;\theta_{{\rm num},i} = \theta} l_{\rm num}(\boldsymbol\theta_{\rm num}).
\]
Note that $\widetilde{l}_{{\rm num},i}(\theta)$
is computed by fixing the $i$-th parameter
to the value $\theta$ 
and minimizing the negative double log-likelihood \eqref{eq:loglikelihood} 
over the remaining parameters.
The $1-\alpha$ confidence interval 
for $\theta_{{\rm num},i}$
derived from the likelihood-ratio test
is given by
\begin{align} \label{eq:proflikelihoodci}
        \big[\theta_{{\rm num},i}^-,\theta_{{\rm num},i}^+\big] = \{\theta : \widetilde{l}_{{\rm num},i}(\theta) -l_{\rm num}\big(\widehat{\boldsymbol\theta}_{\rm num}\big)
        \leq \chi^2_{1,1-\alpha}\},
\end{align}
where $\widehat{\boldsymbol\theta}_{\rm num}$ is the MLE estimator 
defined by \eqref{eq:mleestimator} and $\chi^2_{1,1-\alpha}$ is the $(1-\alpha)$-th quantile of the $\chi^2$-distribution.
Instead of computing the endpoints $\theta_{{\rm num},i}^-$ and $\theta_{{\rm num},i}^+$ directly using \eqref{eq:proflikelihoodci}, they can be computed
by solving 
the following two constrained optimization problems: 
\begin{align} \label{eq:cioptproblem}
\begin{split}
        & \theta_{{\rm num},i}^- = \textstyle \min_{\boldsymbol\theta_{\rm num} \in \boldsymbol\Theta_{\rm num}} \big\{\theta_{{\rm num},i}: l_{\rm num}\big(\boldsymbol\theta_{\rm num}\big) \leq  l_{\rm num} \big(\widehat{\boldsymbol\theta}_{\rm num}\big)+\chi^2_{1,1-\alpha}\big\}, \\
    & \theta_{{\rm num},i}^+ = \textstyle \max_{\boldsymbol\theta_{\rm num} \in \boldsymbol\Theta_{\rm num}} \big\{\theta_{{\rm num},i}: l_{\rm num}\big(\boldsymbol\theta_{\rm num}\big) \leq l_{\rm num}\big(\widehat{\boldsymbol\theta}_{\rm num}\big)+\chi^2_{1,1-\alpha}\big\}.
\end{split}
\end{align}
We refer to e.g.~\cite{neale1997use,fischer2021robust,borisov2020confidence,venzon1988method,raue2009structural} for further details.

Our estimation framework 
is based on solving the optimization problems in \eqref{eq:mleestimator} and \eqref{eq:cioptproblem} using the sqp solver in MATLAB.
The implementation is described in Appendix \ref{app:implementation}.

\subsection{Estimation for cell fraction data} \label{app:estimationcellfrac}

For cell fraction data, we begin by extending the definitions of ${\bf p}^{(j)}(t)$ and ${\bf S}^{(j)}(t)$  from \eqref{eq:QalphaSalphadef} in the main text to a general vector $\boldsymbol\alpha$ of starting cell fractions:
\begin{align} \label{eq:QalphaSalphadefgeneral}
\begin{split}
            &{\bf p}^{\bfalpha}(t) := \big({\bf m}^{\boldsymbol\alpha}(t){\bf 1}^T\big)^{-1} {\bf m}^{\boldsymbol\alpha}(t), \\
        &\textstyle {\bf Q}^{\bfalpha}(t) :=
        {\bf I} - {\bf 1}^T {\bf p}^{\boldsymbol\alpha}(t), \\
&{\bf S}^{\bfalpha}(t) := \big({\bf m}^{\boldsymbol\alpha}(t){\bf 1}^T\big)^{-2}\big({\bf Q}^{\bfalpha}(t)\big)^T \, {\boldsymbol\Sigma}^{{\bfalpha}}(t)  \, {\bf Q}^{\bfalpha}(t).
\end{split}
\end{align}
Then, based on a generalized version of the central limit theorem \eqref{eq:CLTfrac},
which is stated and proved as Proposition \ref{prop:CLTfrac} in Appendix \ref{app:CLT2},
we propose the following extension of the statistical model \eqref{eq:statisticalmodelfrac} in the main text:
\begin{align} \label{eq:statisticalmodelfracgeneral}
    {\bf f}_{i,\ell,r} \; {\sim} \; {\bf p}^{{\bf f}_i}(t_\ell) + {\cal N}\big({\bf 0},N_i^{-1}{\bf S}^{{\bf f}_i}(t_\ell)\big) + {\cal N}\big({\bf 0}, {\bf E}_{i,\ell}^{\rm frac}\big).
\end{align}
Note that the mean behavior ${\bf p}^{{\bf f}_i}(t_\ell)$ depends only on $(\nu_{ik})_{k \neq i}$ and $\boldsymbol\lambda^{[-1]}$, while the variance term $N_i^{-1}{\bf S}^{{\bf f}_i}(t_\ell)$ depends on all model parameters $(\nu_{ik})_{k \neq i},\boldsymbol\lambda^{[-1]},\lambda_1,{\bf d}$.
The choice of type-1 as a reference phenotype is arbitrary,
and we use ${\bf d}$ as opposed to ${\bf b}$ as we found
it to perform well numerically.
As stated in the main text, we assume that the $K \times K$ covariance matrix ${\bf E}_{i,\ell}^{\rm frac}$ associated with measurement error can be written as a function 
of 
${\bf d},\lambda_1,\boldsymbol\lambda^{[-1]},(\nu_{ik})_{k \neq i}$
and added error parameters $\boldsymbol\omega_{\rm frac} = (\omega_{1},\ldots,\omega_{M_{\rm frac}})$
for some $M_{\rm frac} \geq 0$.
We let $\boldsymbol\theta_{\rm frac}$ denote the complete $1 \times \big(K(K+1)+M_{\rm frac}\big)$ 
vector of model parameters including the error parameters.

When deriving a likelihood function 
for the statistical model \eqref{eq:statisticalmodelfracgeneral},
we note that the last coordinate of ${\bf f}_{i,\ell,r}$ provides no new information over the first $K-1$ coordinates, since the coordinates always sum to one.
In the likelihood function, we 
therefore only consider the first $K-1$ coordinates, which we can accomplish by multiplying ${\bf f}_{i,\ell,r}$ by the $K \times (K-1)$ matrix 
${\bf B}$ with 1 on the diagonal and 0 off it.
In this way, we obtain the following likelihood:
    \begin{align} \label{eq:likelihoodfrac}
    &{\cal L}_{\rm frac}\big(\boldsymbol\theta_{\rm frac}\big|({\bf f}_{i,\ell,r})
    \big) \nonumber \\
    &= \textstyle \prod_{i=1}^I \prod_{\ell=1}^{L} \prod_{r=1}^{R} \! \Big((2\pi)^{K-1} {\rm det}\big({\bf B}^T \big( N_i^{-1}{\bf S}^{{\bf f}_i}(t_\ell)+{\bf E}_{i,\ell}^{\rm frac}\big) {\bf B}\big)\Big)^{-1/2} \nonumber \\
    &\quad \textstyle \cdot\exp\Big(\!-\!\frac12 \big({\bf f}_{i,\ell,r}-{\bf p}^{{\bf f}_i}(t_\ell) \big) \,{\bf B}\, \textstyle \big({\bf B}^T\big( N_i^{-1} {\bf S}^{{\bf f}_i}(t_\ell)+{\bf E}_{i,\ell}^{\rm frac}\big)\,{\bf B}\big)^{-1} 
    \, {\bf B}^T \, \big({\bf f}_{i,\ell,r}-{\bf p}^{{\bf f}_i}(t_\ell)\big)^T\Big).
\end{align}
As for cell number data, we define the 
negative double log-likelihood,
\begin{align} \label{eq:loglikelihood2}
    & l_{\rm frac}\big(
    {\boldsymbol\theta}_{\rm frac}
    \big) 
    := -2\log {\cal L}_{\rm frac} \big({\boldsymbol\theta}_{\rm frac}
    \big|({\bf f}_{i,\ell,r})
    \big),
\end{align}
and obtain the maximum likelihood estimate for
${\boldsymbol\theta}_{\rm frac}$ 
by solving
\begin{align} \label{eq:mleestimator2}
    \widehat{\boldsymbol\theta}_{\rm frac}
    := {\rm argmin}_{\boldsymbol\theta_{\rm frac} \in \boldsymbol\Theta_{\rm frac}} \,
    l_{\rm frac}(\boldsymbol\theta_{\rm frac}).
\end{align}
In the feasible set $\boldsymbol\Theta_{\rm frac}$, we restrict the parameter values so that $\nu_{ik} \geq 0$, 
${\bf d} \geq {\bf 0}$, $\lambda_1 \geq -d_1$ and $(\lambda_j-\lambda_1)+d_j+\lambda_1 \geq 0$ for $j=2,\ldots,K$.
Further restrictions can be made depending on the context, see e.g.~Section \ref{sec:realdata} and Appendix \ref{app:implementation}.
The computation of
confidence intervals proceeds as described for cell number data.

For the simplified model \eqref{eq:statisticalmodelfracsimple}, we proceed as above except we remove all terms involving $N_i^{-1} {\bf S}^{{\bf f}_i}(t_\ell)$.

\section{Estimation for reducible switching dynamics} \label{sec:extensions}

\begin{figure}
        \centering
        \includegraphics[scale=1]{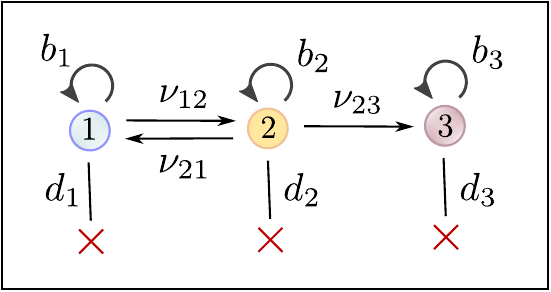}
        \caption[To demonstrate that our estimation framework is applicable to reducible switching models, we consider a three-type model with a reversible transition between type-1 and type-2, and an irreversible transition from type-2 to type-3.]{To demonstrate that our estimation framework is applicable to reducible switching models, we consider a three-type model with a reversible transition between type-1 and type-2, and an irreversible transition from type-2 to type-3. This model is applicable e.g.~to epigenetic gene silencing under the recruitment of chromatin regulators \cite{bintu2016dynamics} and to epigenetically-driven drug resistance in cancer \cite{gunnarsson2020understanding}.}
        \label{fig:extensions}
    \end{figure}

In the main text, we have assumed that the switching dynamics are irreducible,
meaning that it is possible to switch between any pair of phenotypes,
possibly through intermediate types.
In this section, we show how our framework can be applied to the case of reducible switching dynamics.
For simplicity, we will consider one particular model shown in Figure \ref{fig:extensions}.
This model has been applied e.g.~to the dynamics of epigenetic gene silencing under recruitment of chromatin regulators \cite{bintu2016dynamics}
and the evolution of epigenetically-driven drug resistance in cancer,
where drug-sensitive cells (type-1) first acquire a transiently resistant phenotype (type-2) and then evolve to stable epigenetic resistance (type-3) \cite{gunnarsson2020understanding}.

Say that experiments are conducted from isolated initial conditions,
and say first that cell number data is collected.
For the model in Figure \ref{fig:extensions}, the distribution of the data vector ${\bf n}_{3,\ell,r}$ is degenerate,
since $n_{3,\ell,r,j} = 0$ for $j = 1,2$.
As a result, the covariance matrix $\boldsymbol\Sigma^{(3)}(t_\ell)$ is singular for all $\ell=1,\ldots,L$, and the likelihood function in \eqref{eq:likelihoodnum} is not defined.
To resolve this issue, we set ${\bf C}_1 = {\bf C}_2 = {\bf I}$ and ${\bf C}_3 = {\bf e}_3^T$,
where ${\bf e}_3$ is the $1 \times K$ third unit vector.
By Proposition \ref{prop:CLTnum}, ${\bf n}_{3,\ell,r}{\bf C}_3 = n_{3,\ell,r,3}$ has a normal distribution, which is nondegenerate.
We therefore modify the likelihood function in \eqref{eq:likelihoodnum} to
\begin{linenomath*}
    \begin{align*}
    &{\cal L}_{\rm num}\big(
    {\boldsymbol\theta}_{\rm num}
    \big|({\bf n}_{i,\ell,r})_{i,\ell,r}\big) \nonumber \\
    &= \textstyle \prod_{i=1}^3 \prod_{\ell=1}^{L} \prod_{r=1}^{R} \! \Big((2\pi)^{K} {\rm det}\big({\bf C}_i^T \big( N_i{\boldsymbol\Sigma}^{(i)}(t_\ell)+{\bf E}_{i,\ell}^{\rm num}\big) {\bf C}_i \big)\Big)^{-1/2} \nonumber \\
    &\quad \textstyle \cdot\exp\big(\!-\!\frac12 \big({\bf n}_{i,\ell,r}-N_i{\bf m}^{(i)}(t_\ell)\big) {\bf C}_i  \big({\bf C}_i^T \big( N_i{\boldsymbol\Sigma}^{(i)}(t_\ell)+{\bf E}_{i,\ell}^{\rm num}\big) {\bf C}_i \big)^{-1} {\bf C}_i^T \big({\bf n}_{i,\ell,r}-N_i{\bf m}^{(i)}(t_\ell)\big)^T\big).
\end{align*}
\end{linenomath*}
From this likelihood function, MLE estimates
and confidence intervals can be computed as described in Appendix \ref{app:estimationdeets}, where we restrict the set of feasible parameters $\boldsymbol\Theta_{\rm num}$ so that $\nu_{13} = \nu_{31} = \nu_{32} = 0$.
By our analysis in Section \ref{sec:identifiabilitynum},
all model parameters are structurally identifiable for this example.

To accommodate model structures such as the one discussed here,
the above modified likelihood function is implemented in our MATLAB codes (Appendix \ref{app:implementation}).
By taking ${\bf C}_i = {\bf I}$ for each $i=1,\ldots,I$,
we recover the original likelihood function in \eqref{eq:likelihoodnum}.

If cell fraction data is collected, there is no value in conducting experiments starting only from type-3 cells.
We therefore use the likelihood function
\begin{linenomath*}
\begin{align*} 
    &{\cal L}_{\rm frac}\big(\boldsymbol\theta_{\rm frac}\big|({\bf f}_{i,\ell,r})_{i,\ell,r}\big) \nonumber \\
    &= \textstyle \prod_{i=1}^2 \prod_{\ell=1}^{L} \prod_{r=1}^{R} \! \Big((2\pi)^{K-1} {\rm det}\big({\bf B}^T \big( N_i^{-1}{\bf S}^{(i)}(t_\ell)+{\bf E}_{i,\ell}^{\rm frac}\big) {\bf B}\big)\Big)^{-1/2} \nonumber \\
    &\quad \textstyle \cdot\exp\Big(\!-\!\frac12 \big({\bf f}_{i,\ell,r}-{\bf p}^{(i)}(t_\ell) \big){\bf B}  \textstyle \big({\bf B}^T\big( N_i^{-1} {\bf S}^{(i)}(t_\ell)+{\bf E}_{i,\ell}^{\rm frac}\big)\,{\bf B}\big)^{-1} 
    {\bf B}^T \big({\bf f}_{i,\ell,r}-{\bf p}^{(i)}(t_\ell)\big)^T\Big),
\end{align*}
\end{linenomath*}
where we only include experiments started by type-1 and type-2 cells, respectively.
By our analysis in Section \ref{sec:identifiabilitycellfraction},
the switching rates $\nu_{12}$, $\nu_{21}$ and $\nu_{23}$, and the net birth rate differences $\lambda_2-\lambda_1$ and $\lambda_3-\lambda_2$, are structurally identifiable in this case.
An example of a model structure where it becomes necessary to modify the above likelihood function for cell fraction data is given in Appendix \ref{app:implementation}.

\section{Central limit theorems} \label{app:CLTs}

In the main text, we stated the central limit theorems (CLTs) \eqref{eq:CLTnum} and \eqref{eq:CLTfrac} for cell number and cell fraction data, respectively, for the simple case where all experiments are started from isolated subpopulations.
Here, we state and prove the CLTs for general starting conditions.

\subsection{Cell number data} \label{app:CLT1}

For cell number data,
we begin by modifying the notation developed for the branching process model in Section \ref{sec:randomproc} to facilitate analysis of general starting conditions.
In particular, for a general $1 \times K$ vector ${\bf n}=(n_1,\ldots,n_K)$ of starting cell numbers of each type, we let ${\bf Z}^{{\bf n}}(t) = \big(Z^{{\bf n}}_1(t),\ldots,Z^{{\bf n}}_K(t)\big)$ denote the random vector of cell numbers at time $t$.
We state and prove a CLT for ${\bf Z}^{{\bf n}}(t)$ when the total number of starting cells is sent to infinity (Proposition \ref{prop:CLTnum}).
More precisely, we fix a vector ${\boldsymbol\alpha}$ of starting cell fractions with $\alpha_i \geq 0$ for $i=1,\ldots,K$ and $\sum_{i=1}^K \alpha_i=1$, 
write the vector of starting cell numbers as ${\bf n} = \lfloor N\boldsymbol\alpha\rfloor$,
and send $N \to \infty$.
Note that some coordinates of the vector ${\boldsymbol\alpha}$ are allowed to be 0.
In the $N \to \infty$ regime, the starting condition ${\bf n} = \lfloor N{\boldsymbol\alpha}\rfloor$ will therefore either include no cell or a large number of cells of any given type.
This is consistent with our assumptions on the vectors ${\bf n}_1,\dots,{\bf n}_I$ of experimental starting conditions (Appendix \ref{app:experimentsanddata}).

We establish Proposition \ref{prop:CLTnum} 
more generally
for linear transformations
${\bf Z}^{{\bf n}}(t)\, {\bf C}$
of ${\bf Z}^{{\bf n}}(t)$,
which gives a CLT for ${\bf Z}^{\bf n}(t)$ by taking ${\bf C} = {\bf I}$.
The more general version
allows us to obtain a CLT 
for cases where we
do not observe the
full vector ${\bf Z}^{{\bf n}}(t)$.
For example, if we 
set ${\bf C} := {\bf 1}^T$, then ${\bf Z}^{{\bf n}}(t)\, {\bf C} = \sum_{k=1}^K Z_k^{{\bf n}}(t)$
is the total
number of cells at time $t$.
The more general version also becomes useful when estimating from models with reducible switching dynamics, as we discussed in Appendix \ref{sec:extensions}.

\begin{proposition} \label{prop:CLTnum}
Let $\boldsymbol\alpha$ be $1 \times K$ with $\alpha_i \geq 0$ for $i=1,\ldots,K$ and $\sum_{i=1}^K \alpha_i=1$.
Let $J \geq 1$ be any integer. For any $K \times J$ matrix ${\bf C}$, then as $N \to \infty$,
\[
        N^{-1/2}\big({\bf Z}^{\lfloor N\bfalpha\rfloor}(t)\, {\bf C}-N{\bf m}^{\bfalpha}(t)\,{\bf C}\big) \stackrel{d}{\to} {\cal N}\big({\bf 0},{\bf C}^T\bfSigma^{\bfalpha}(t)\,{\bf C}\big).
\]
Here, 
the covariance matrix
${\boldsymbol\Sigma}^{(j)}(t)$ is given by
\begin{linenomath*}
        \begin{align*} 
    \begin{split}
                \textstyle {\boldsymbol\Sigma}^{(j)}(t) &= \textstyle  2 \int_0^t ({\bf M}(t-\tau))^T {\rm diag}\big({\bf b} \odot {\bf m}^{(j)}(\tau)\big) ({\bf M}(t-\tau))d\tau \\
        &\quad \textstyle +{\rm diag}\big({\bf m}^{(j)}(t)\big)- ({\bf m}^{(j)}(t))^T{\bf m}^{(j)}(t).
    \end{split}
    \end{align*}
\end{linenomath*}
\end{proposition}

\begin{proof} 
First note that we can write
  \begin{align} \label{eq:indsum}
        \textstyle {\bf Z}^{\lfloor N\bfalpha\rfloor}(t) = \sum_{j=1}^K {\bf Z}^{\lfloor N\alpha_j\rfloor {\bf e}_j}(t),
    \end{align}
    where $\big({\bf Z}^{\lfloor N\alpha_j\rfloor {\bf e}_j}(s)\big)_{s \geq 0}$ for $j=1,\ldots,K$ are independent branching processes started with $\lfloor N\alpha_j\rfloor$ cells of type-$j$, respectively.
    For each process, we can write
       \[
        \textstyle {\bf Z}^{\lfloor N\alpha_j\rfloor {\bf e}_j}(t) = \sum_{m=1}^{\lfloor N\alpha_j\rfloor} {\bf Z}^{(j),m}(t),
    \]
    where $\big({\bf Z}^{(j),m}(s)\big)_{s \geq 0}$ for $m=1,\ldots,\lfloor N\alpha_j\rfloor$ are i.i.d.~copies of the branching process $\big({\bf Z}^{(j)}(s)\big)_{s \geq 0}$ started by a single type-$j$ cell.
    Set
    \begin{align} \label{eq:Wdef}
    & \textstyle {\bf W}^{\lfloor N\alpha_j\rfloor{\bf e}_j}(t) :=   N^{-1/2} \big({\bf Z}^{\lfloor N\alpha_j\rfloor {\bf e}_j}(t)- N \alpha_j {\bf m}^{(j)}(t)\big).
\end{align}
Let $J \geq 1$ and let ${\bf C}$ be a $K \times J$ matrix.
    By the standard (multivariate) central limit theorem, as $N \to \infty$,
    \[
      {\bf W}^{\lfloor N\alpha_j\rfloor{\bf e}_j}(t)\,{\bf C} \stackrel{d}{\to} {\cal N}\big({\bf 0},\alpha_j {\bf C}^T\bfSigma^{(j)}(t)\,{\bf C}\big),
        \]
        where ${\boldsymbol\Sigma}^{(j)}(t)$ is the covariance matrix for ${\bf Z}^{(j)}(t)$.
        We can then conclude from \eqref{eq:indsum} that as $N \to \infty$,
        \[
            N^{-1/2}\big({\bf Z}^{\lfloor N\alpha\rfloor}(t)\,{\bf C}-N{\bf m}^{\bfalpha}(t)\,{\bf C}\big) \stackrel{d}{\to} {\cal N}\big({\bf 0},{\bf C}^T\bfSigma^{\bfalpha}(t)\,{\bf C}\big).
       \]
        It remains to derive the given expression 
        for the covariance matrix ${\boldsymbol\Sigma}^{(j)}(t)$.
        To that end,
let ${\bf D}^{(j)}(t)$ be the matrix of second factorial moments of ${\bf Z}^{(j)}(t)$,
    \[
        D_{k\ell}^{(j)}(t) := \E\big[Z_k^{(j)}(t)\big(Z_\ell^{(j)}(t)-\delta_{k\ell}\big)\big],
    \]
    where $\delta_{k\ell}$ is the Kronecker delta.
    Let ${\bf s}=(s_1,\ldots,s_K)$ be a $K$-dimensional vector of real numbers and set $h_j := b_j+d_j+\sum_{k \neq j} \nu_{jk}$ for $j=1,\ldots,K$.
    Furthermore, let
    \[
            u^{(j)}({\bf s}) := b_j s_j^2 + d_j + \sum_{k \neq j} \nu_{jk} s_k - h_j s_j, \quad  {\bf 0} \leq {\bf s} \leq {\bf 1},
    \]
    be the infinitesimal generating function for ${\bf Z}^{(j)}(t)$, and let
    \[
        \textstyle F^{(j)}({\bf s},t) := \E\big[{\bf s}^{{\bf Z}^{(j)}(t)}\big] = \E\Big[\prod_{k=1}^K s_k^{Z^{(j)}_k(t)}\Big], \quad {\bf 0} \leq {\bf s} \leq {\bf 1}, \; t \geq 0,
    \]
    be the probability generating function for ${\bf Z}^{(j)}(t)$.
    With this notation, we can write the Kolmogorov forward equation for ${\bf Z}^{(j)}(t)$ as
    \[
        \textstyle \frac{\partial}{\partial t} F^{(j)}({\bf s},t) = \sum_{i=1}^K u^{(i)}({\bf s}) \, \frac{\partial}{\partial s_i} F^{(j)}({\bf s},t).
    \]
    Then, for $k,\ell=1,\ldots,K$,
        \begin{align} \label{kolmbackwdiff}
        &\textstyle \frac{\partial}{\partial t} \big(\frac{\partial}{\partial s_k} \frac{\partial}{\partial s_\ell}  F^{(j)}({\bf s},t)\big) \nonumber \\
        & = \textstyle \sum_{i=1}^K \textstyle \big(\frac{\partial}{\partial s_k}\frac{\partial}{\partial s_\ell} u^{(i)}({\bf s}) \frac{\partial}{\partial s_i} F^{(j)}({\bf s},t) 
        + \frac{\partial}{\partial s_\ell} u^{(i)}({\bf s}) \frac{\partial}{\partial s_k} \frac{\partial}{\partial s_i} F^{(j)}({\bf s},t) \nonumber \\
        &\qquad\qquad\; +  \textstyle \frac{\partial}{\partial s_k} u^{(i)}({\bf s}) \frac{\partial}{\partial s_\ell} \frac{\partial}{\partial s_i} F^{(j)}({\bf s},t) +  u^{(i)}({\bf s}) \frac{\partial}{\partial s_k} \frac{\partial}{\partial s_\ell} \frac{\partial}{\partial s_i} F^{(j)}({\bf s},t)\big).
    \end{align}
    Now,
    \begin{linenomath*}
    \begin{align*}
        &\textstyle \frac{\partial}{\partial s_k} u^{(i)}({\bf s}) = \begin{cases} 2b_is_i-h_i, & k=i, \\ \nu_{ik}, & k \neq i, \end{cases} \\
        &\textstyle \frac{\partial}{\partial s_k}\frac{\partial}{\partial s_\ell} u^{(i)}({\bf s}) = \delta_{ki} \delta_{\ell i} 2b_i.
    \end{align*}
    \end{linenomath*}
Let ${\bf A}$ be the infinitesimal generator and ${\bf M}(t)$ be the mean matrix as defined in Sections \ref{sec:modelparameters} and \ref{sec:randomproc}.
    Since
     \begin{linenomath*}
    \begin{align*}
        &\textstyle a_{ik} = \frac{\partial}{\partial s_k} u^{(i)}({\bf s})|_{{\bf s}={\bf 1}}, \\
        &M_{jk}(t) = \textstyle \frac{\partial}{\partial s_k} F^{(j)}({\bf s},t)\big|_{{\bf s}=1}, \\
        &D_{k\ell}^{(j)}(t) = \textstyle \frac{\partial}{\partial s_k} \frac{\partial}{\partial s_\ell}  F^{(j)}({\bf s},t)\big|_{{\bf s}=1},
    \end{align*}
     \end{linenomath*}
    and $u^{(i)}({\bf 1}) = 0$, we can conclude from \eqref{kolmbackwdiff} that
     \begin{linenomath*}
    \begin{align*}
        \textstyle \frac{d}{d t} D_{k\ell}^{(j)}(t) &= \textstyle \sum_{i=1}^K \big(\delta_{ki}\delta_{\ell i} 2b_i M_{ji}(t)+a_{i\ell} D_{ki}^{(j)}(t)+a_{ik} D^{(j)}_{\ell i}(t)\big) \\
        &= \textstyle \sum_{i=1}^K a_{ik} D^{(j)}_{i \ell}(t)+ \sum_{i=1}^K D_{ki}^{(j)}(t)a_{i\ell} + \delta_{k\ell} 2b_k M_{jk}(t).
    \end{align*}
     \end{linenomath*}
    In the second step, we use that $D_{i\ell}^{(j)}(t) = D_{\ell i}^{(j)}(t)$.
    This yields a Lyapunov matrix differential equation,
    \begin{align} \label{eq:matrixdiffeq}
        \textstyle \frac{d}{d t} {\bf D}^{(j)}(t) = {\bf A}^T{\bf D}^{(j)}(t) + {\bf D}^{(j)}(t){\bf A} + 2\, {\rm diag}\big({\bf b} \odot {\bf m}^{(j)}(t)\big),
    \end{align}
    with initial condition ${\bf D}^{(j)}(0) = {\bf 0}$.
    The solution is
     \begin{linenomath*}
\begin{align*}
    \textstyle {\bf D}^{(j)}(t) &= \textstyle 2 \exp(t{\bf A}^T) \big(\int_0^t \exp(-\tau{\bf A}^T) {\rm diag}\big({\bf b} \odot {\bf m}^{(j)}(\tau)\big) \exp(-\tau {\bf A}) d\tau\big) \exp(t{\bf A}) \\
    &= \textstyle 2 \int_0^t ({\bf M}(t-\tau))^T {\rm diag}\big({\bf b} \odot {\bf m}^{(j)}(\tau)\big) ({\bf M}(t-\tau))d\tau,
\end{align*}
 \end{linenomath*}
and the expression \eqref{eq:covmatrixexpression} for ${\boldsymbol\Sigma}^{(j)}(t)$ follows from the fact that
\begin{align} \label{eq:SigmavsD}
    {\boldsymbol\Sigma}^{(j)}(t) = {\bf D}^{(j)}(t) +{\rm diag}\big({\bf m}^{(j)}(t)\big)- ({\bf m}^{(j)}(t))^T{\bf m}^{(j)}(t).
\end{align}
\end{proof}

\subsection{Cell fraction data} \label{app:CLT2}

For cell fraction data, we similarly begin by modifying the notation developed for the branching process model in Section \ref{sec:randomproc}.
In particular, for the $1 \times K$ vector ${\bf n}=(n_1,\ldots,n_K)$ of starting cell numbers of each type, then on the event $\big\{\sum_{k=1}^K Z_k^{\bf n}(t) \neq 0\big\}$, we 
let $\boldsymbol\Delta^{\bf n}(t)$ denote the random vector of cell fractions at time $t$, i.e.
\[
\textstyle \Delta_i^{\bf n}(t) := Z_i^{\bf n}(t)/\big(\sum_{k=1}^K Z_k^{\bf n}(t)\big), \quad i=1,\ldots,K.
\]
We now state and prove a central limit theorem for ${\boldsymbol\Delta}^{{\bf n}}(t)$ (Proposition \ref{prop:CLTfrac}).
As for cell number data,
the CLT is established for linear transformations ${\boldsymbol\Delta}^{{\bf n}}(t) \, {\bf C}$ of ${\boldsymbol\Delta}^{{\bf n}}(t)$.
We note that the CLT has already been established 
for the case of an isolated large starting population 
by Yakovlev and Yanev 
\cite{yakovlev2009relative}.
We extend their argument to more general starting conditions
by fixing the vector $\boldsymbol\alpha$ of starting cell proportions and sending the total population size $N$ to infinity.
We
also provide a 
simplified
expression for the covariance matrix ${\bf S}^{\bfalpha}(t)$
and 
show that the mean function ${\bf p}^{\boldsymbol\alpha}(t)$ can be written
solely in terms of $(\nu_{ik})_{k \neq i}$ and $\boldsymbol\lambda^{[-1]}$.

\begin{proposition} \label{prop:CLTfrac}
Let 
$\boldsymbol\alpha$ be
$1 \times K$ 
with $\alpha_i \geq 0$ for $i=1,\ldots,K$ and $\sum_{i=1}^K \alpha_i=1$.
Let $J \geq 1$ be any integer. For any $K \times J$ matrix ${\bf C}$, then as $N \to \infty$,
    \[
     N^{1/2}\big(\boldsymbol\Delta^{\lfloor N\bfalpha\rfloor}(t)\,{\bf C}-{\bf p}^{\bfalpha}(t)\,{\bf C}\big) \stackrel{d}{\to} {\cal N}\big({\bf 0},{\bf C}^T{\bf S}^{\bfalpha}(t)\,{\bf C}\big).
    \]
Here, the mean function ${\bf p}^{\boldsymbol\alpha}(t)$ can be written solely as a function of the switching rates $(\nu_{ik})_{k \neq i}$ and the relative net birth rates $\boldsymbol\lambda^{[-1]}$.
\end{proposition}

\begin{proof} 

Recall from \eqref{eq:indsum} 
that we can write
  \[
        \textstyle {\bf Z}^{\lfloor N\bfalpha\rfloor}(t) = \sum_{j=1}^K {\bf Z}^{\lfloor N\alpha_j\rfloor {\bf e}_j}(t),
    \]
    where $\big({\bf Z}^{\lfloor N\alpha_j\rfloor {\bf e}_j}(s)\big)_{s \geq 0}$ for $j=1,\ldots,K$ are independent branching processes started with $\lfloor N\alpha_j\rfloor$ cells of type-$j$, respectively.
Define
\[
\textstyle U^{\lfloor N{\boldsymbol\alpha}\rfloor}(t) := \sum_{k=1}^K Z_k^{\lfloor N{\boldsymbol\alpha}\rfloor}(t) = \sum_{j=1}^K \sum_{k=1}^K Z_k^{\lfloor N \alpha_j\rfloor {\bf e}_j}(t)
\]
as the total population size at time $t$ and note that
\[
    \textstyle \Delta_i^{\lfloor N{\boldsymbol\alpha}\rfloor}(t) = \frac{Z_i^{\lfloor N{\boldsymbol\alpha}\rfloor}(t)}{U^{\lfloor N{\boldsymbol\alpha}\rfloor}(t)} = \frac{\sum_{j=1}^K Z_i^{\lfloor N\alpha_j\rfloor {\bf e}_j}(t)}{U^{\lfloor N{\boldsymbol\alpha}\rfloor}(t)}.
\]
We can therefore write
 \begin{linenomath*}
\begin{align*}
    & \sqrt{N}\big(\Delta_i^{\lfloor N{\boldsymbol\alpha}\rfloor}(t)-p_i^{\boldsymbol\alpha}(t)\big) \\
    &= \textstyle \frac{\sqrt{N}}{U^{\lfloor N{\boldsymbol\alpha}\rfloor}(t)}\Big( \sum_{j=1}^K \! \Big( (1-p_i^{\boldsymbol\alpha}(t)) Z_i^{\lfloor N \alpha_j\rfloor {\bf e}_j}(t) - p_i^{\boldsymbol\alpha}(t) \sum_{k \neq i} Z_k^{\lfloor N \alpha_j\rfloor {\bf e}_j}(t) \Big) \Big).
\end{align*}
 \end{linenomath*}
Note that by definition,
\[
    \textstyle {\bf p}^{\bfalpha}(t) = \big({\bf m}^{\boldsymbol\alpha}(t){\bf 1}^T\big)^{-1} {\bf m}^{\boldsymbol\alpha}(t) = \Big(\sum_{j=1}^K\sum_{k=1}^K \alpha_j m_k^{(j)}(t)\Big)^{-1} \sum_{j=1}^K \alpha_j {\bf m}^{(j)}(t).
\]
It follows that
 \begin{linenomath*}
\begin{align*}
    & \textstyle \sum_{j=1}^K N \alpha_j \Big((1-p_i^{\boldsymbol\alpha}(t)) m_i^{(j)}(t) - p_i^{\boldsymbol\alpha}(t) \sum_{k \neq i} m_k^{(j)}(t)\Big) \\
    & = \textstyle N \Big(\sum_{j=1}^K \alpha_j m_i^{(j)}(t) - p_i^{\boldsymbol\alpha}(t) \sum_{j=1}^K \sum_{k=1}^K \alpha_j m_k^{(j)}(t)\Big) \\
    &= 0.
\end{align*}
 \end{linenomath*}
We can therefore write
 \begin{linenomath*}
\begin{align*}
    & \sqrt{N}\big(\Delta_i^{\lfloor N{\boldsymbol\alpha}\rfloor}(t)-p_i^{\boldsymbol\alpha}(t)\big) \\
    &= \textstyle \frac{N}{U^{\lfloor N{\boldsymbol\alpha}\rfloor}(t)}\Big( \sum_{j=1}^K \Big( (1-p_i^{\boldsymbol\alpha}(t)) W_i^{\lfloor N \alpha_j\rfloor {\bf e}_j}(t) - p_i^{\boldsymbol\alpha}(t) \sum_{k \neq i} W_k^{\lfloor N \alpha_j\rfloor {\bf e}_j}(t) \Big) \Big),
\end{align*}
 \end{linenomath*}
where the vector ${\bf W}^{\lfloor N\alpha_j\rfloor {\bf e}_j}(t)$ is defined as in \eqref{eq:Wdef}.
In vector form, this becomes
\[
    \textstyle \sqrt{N} \big({\boldsymbol\Delta}^{\lfloor N\bfalpha\rfloor}(t)-{\bf p}^{\bfalpha}(t)\big) = \frac{N}{U^{\lfloor N{\boldsymbol\alpha}\rfloor}(t)} \sum_{j=1}^K {\bf W}^{\lfloor N\alpha_j\rfloor{\bf e}_j}(t){\bf Q}^{\bfalpha}(t),
\]
where ${\bf Q}^{\boldsymbol\alpha}(t)$ is defined as in \eqref{eq:QalphaSalphadefgeneral}.
By the strong law of large numbers, $U^{\lfloor N{\boldsymbol\alpha}\rfloor}(t)/N \to {\bf m}^{\boldsymbol\alpha}(t){\bf 1}^T$ almost surely as $N \to \infty$.
Let $J \geq 1$ and let ${\bf C}$ be a $K \times J$ matrix.
By the standard (multivariate) central limit theorem, as $N \to \infty$,
\[
{\bf W}^{\lfloor N\alpha_j\rfloor{\bf e}_j}(t)\,{\bf Q}^{\bfalpha}(t)\,{\bf C} \stackrel{d}{\to} {\cal N}\Big({\bf 0},\alpha_j{\bf C}^T\big({\bf Q}^{\bfalpha}(t)\big)^T{\bfSigma}^{(j)}(t){\bf Q}^{\bfalpha}(t)\,{\bf C}\Big).
\]
Writing ${\boldsymbol\Sigma}^{\bfalpha}(t) = \sum_{j=1}^K \alpha_j {\boldsymbol\Sigma}^{(j)}(t)$,
it finally follows from 
Slutsky's theorem that
\[
    \textstyle \sqrt{N} \big({\boldsymbol\Delta}^{\lfloor N\bfalpha\rfloor}(t)\,{\bf C}-{\bf p}^{\bfalpha}(t)\,{\bf C}\big) \stackrel{d}{\to} {\cal N}\Big({\bf 0},\big({\bf m}^{\bfalpha}(t){\bf 1}^T\big)^{-2}{\bf C}^T\big({\bf Q}^{\bfalpha}(t)\big)^T{\bfSigma}^{\boldsymbol\alpha}(t){\bf Q}^{\bfalpha}(t)\,{\bf C}\Big). \qquad 
\]
It remains to show that ${\bf p}^{\bfalpha}(t)$ can be written solely as a function of the switching rates $(\nu_{ik})_{k \neq i}$ and the net birth rate differences $\boldsymbol\lambda^{[-1]}$.
To this end, we define for any $j=1,\ldots,K$:
\begin{align} \label{eq:A-jdef}
\begin{split}
        & {\bf A}^{[-j]} := {\bf A}-\lambda_j {\bf I},
\end{split}
\end{align}
where ${\bf I}$ is the $K \times K$ identity matrix,
and
\begin{align} \label{eq:M-jdef}
\begin{split}
    & \textstyle {\bf M}^{[-j]}(t) := \exp\big(t{\bf A}^{[-j]}\big) = \sum_{k=0}^\infty (t^k/k!) \big({\bf A}^{[-j]}\big)^k, \quad t \geq 0.
\end{split}
\end{align}
Note that ${\bf A}^{[-j]}$ and ${\bf M}^{[-j]}(t)$ only depend on 
$(\nu_{ik})_{k \neq i}$ and $\boldsymbol\lambda^{[-j]}$.
It is easy to see that
\[
    {\bf M}(t) = e^{\lambda_j t} {\bf M}^{[-j]}(t),
\]
for $j=1,\ldots,K$, from which it follows that
\begin{align}  \label{eq:palphanetbirthrates}
\textstyle {\bf p}^{\bfalpha}(t) 
& = \big({\boldsymbol\alpha {\bf M}(t) {\bf 1}^T}\big)^{-1} \boldsymbol\alpha {\bf M}(t) \nonumber \\
& = \big(\bfalpha {\bf M}^{[-1]}(t){\bf 1}^T\big)^{-1} \big({\bfalpha {\bf M}^{[-1]}(t)}\big), \quad t \geq 0.
\end{align}
This completes the proof.
\end{proof}

\section{Proof of Proposition \ref{cor:idcellnum}} \label{app:identcellnum}

\begin{proof}[Proof of Proposition \ref{cor:idcellnum}]
\begin{enumerate}[(1)]
    \item Since ${\bf M}(t) = \exp(t{\bf A}) = \sum_{k=0}^\infty (1/k!)t^k{\bf A}^k$, we have $\frac{d}{dt} {\bf M}(t) = {\bf A} {\bf M}(t)$.
    By taking $t=0$ and noting that ${\bf M}(0) = {\bf I}$, we obtain
    \[
        \textstyle \frac{d}{dt} {\bf M}(t) \big|_{t = 0} = {\bf A}.
    \]
    If $\frac{d}{dt} {\bf M}(t) \big|_{t = 0}$ is known, we can recover the switching rate $\nu_{jk}$ for $k \neq j$ by recalling that $a_{jk} = \nu_{jk}$.
    We can then recover $\lambda_j$ for $j=1,\ldots,K$ by recalling that $a_{jj} = \lambda_j - \sum_{k \neq j} \nu_{jk}$.
    \item Recall that ${\bf m}^{(j)}(t) = {\bf e}_j{\bf M}(t)$. By \eqref{eq:SigmavsD} in the proof of Proposition \ref{prop:CLTnum}, we can write
     \begin{linenomath*}
\begin{align*}
    &\textstyle \frac{d}{dt} {\boldsymbol\Sigma}^{(j)}(t) \\
    &= \textstyle \frac{d}{dt} {\bf D}^{(j)}(t) + {\rm diag}\big({\bf e}_j{\bf A}{\bf M}(t)\big) - {\bf A}^T\big({\bf M}(t)\big)^T{\bf e}_j^T{\bf e}_j{\bf M}(t) - \big({\bf M}(t)\big)^T{\bf e}_j^T{\bf e}_j{\bf A}{\bf M}(t),
\end{align*}
 \end{linenomath*}
where ${\bf D}^{(j)}(t)$ is the matrix of second factorial moments of ${\bf Z}^{(j)}(t)$.
Next, by taking $t=0$ in \eqref{eq:matrixdiffeq} and noting that ${\bf D}^{(j)}(0) = {\bf 0}$ and ${\bf m}^{(j)}(0) = {\bf e}_j$ for all $j=1,\ldots,K$, we see that
\[
   \textstyle \frac{d}{dt} {\bf D}^{(j)}(t) \big|_{t = 0} = 2b_j{\bf e}_j^T{\bf e}_j. 
\]
It follows that
\begin{align} \label{eq:ddtSigma}
    \textstyle \frac{d}{dt} {\boldsymbol\Sigma}^{(j)}(t) \big|_{t = 0} = 2b_j {\bf e}_j^T{\bf e}_j + {\rm diag}\big({\bf e}_j{\bf A}\big) - \big({\bf e}_j^T{\bf e}_j{\bf A}\big)^T - {\bf e}_j^T{\bf e}_j{\bf A}.
\end{align}
For each $j=1,\ldots,K$, if the switching rates $\nu_{jk}$ for $k \neq j$ and the net birth rate $\lambda_j$ are known, the birth rate $b_j$ can be recovered from $\big(\frac{d}{dt} {\boldsymbol\Sigma}^{(j)}(t) \big|_{t = 0}\big)_{jj}$ using this expression.
\qedhere
\end{enumerate}
\end{proof}

\section{Proof of Proposition \ref{cor:idcellfrac1}} \label{app:fracident1}

\begin{proof}[Proof of Proposition \ref{cor:idcellfrac1}]
We begin by establishing some notation.
First, define
${\bf Q}^{(j)}(t) := {\bf Q}^{{\bf e}_j}(t)$
and 
${\bf Q}^{(j)} := {\bf Q}^{(j)}(0) = {\bf I}-{\bf 1}^T{\bf e}_j$,
with ${\bf Q}^{{\bf e}_j}(t)$ defined as in \eqref{eq:QalphaSalphadefgeneral}.
Also define
\begin{align} \label{eq:Vdef}
    {\bf V} := {\bf A} - {\rm diag}\big(\boldsymbol\lambda\big)
\end{align}
as the infinitesimal generator ${\bf A}$ with the net birth rates $\boldsymbol\lambda$ removed from the diagonal.
Let ${\bf v}^{(j)}$ denote the $j$-th row vector of ${\bf V}$
with coordinates $v^{(j)}_k = \nu_{jk}$ for $k \neq j$ and $v^{(j)}_j = -\sum_{k \neq j} \nu_{jk}$, and note that
\begin{align} \label{eq:V-jdef}
    {\bf v}^{(j)} = {\bf e}_j{\bf V} = {\bf e}_j{\bf A}^{[-j]},
\end{align}
where ${\bf A}^{[-j]}$ is defined as in \eqref{eq:A-jdef}.
Also note that ${\bf v}^{(j)} {\bf 1}^T = 0$.
In the proof, we will rely on the following basic facts:
\begin{align} \label{eq:ejQjzero}
\begin{split}
        & {\bf e}_j{\bf Q}^{(j)} = {\bf e}_j\big({\bf I}-{\bf 1}^T{\bf e}_j\big) = {\bf 0}, \\
    & {\bf v}^{(j)} {\bf Q}^{(j)} = {\bf v}^{(j)} \big({\bf I}-{\bf 1}^T{\bf e}_j\big) = {\bf v}^{(j)}.
\end{split}
\end{align}
\begin{enumerate}[(1)]
    \item Since ${\bf p}^{(j)}(t) = \big({\bf e}_j\exp(t{\bf A}){\bf 1}^T\big)^{-1}\big({\bf e}_j\exp(t{\bf A})\big)$, we can write
\begin{align} \label{eq:pfirstder}
\begin{split}
        \textstyle \frac{d}{dt} {\bf p}^{(j)}(t) &= \big({\bf e}_j\exp(t{\bf A}){\bf 1}^T\big)^{-1} \big({\bf e}_j {\bf A} \exp(t{\bf A})\big) \\
    &\quad -\big({\bf e}_j\exp(t{\bf A}){\bf 1}^T\big)^{-2}\big({\bf e}_j{\bf A}\exp(t{\bf A}){\bf 1}^T\big) \big({\bf e}_j \exp(t{\bf A})\big).
\end{split}
\end{align}
Since $\exp({\bf 0}) = {\bf I}$, ${\bf e}_j{\bf 1}^T=1$ and ${\bf e}_j{\bf A}{\bf 1}^T = \lambda_j$, we obtain by \eqref{eq:V-jdef},
\begin{align} \label{eq:meanfirstder}
    \textstyle \frac{d}{dt} {\bf p}^{(j)}(t)\big|_{t = 0} &
    = {\bf e}_j({\bf A}-\lambda_j{\bf I}) 
    = {\bf e}_j{\bf A}^{[-j]}
    = {\bf v}^{(j)}.
 \end{align}
Since the $k$-th coordinate of ${\bf v}^{(j)}$ is $\nu_{jk}$ for $k \neq j$, we can recover $\nu_{jk}$ from the $k$-th coordinate of $\frac{d}{dt} {\bf p}^{(j)}(t)\big|_{t = 0}$.
\item 
\begin{enumerate}[(i)]
    \item Using \eqref{eq:pfirstder}, we begin by writing
     \begin{linenomath*}
\begin{align*}
    \textstyle \frac{d^2}{dt^2} {\bf p}^{(j)}(t) &= \big({\bf e}_j\exp(t{\bf A}){\bf 1}^T\big)^{-1} \big({\bf e}_j {\bf A}^2 \exp(t{\bf A})\big) \\
    &\quad -\big({\bf e}_j\exp(t{\bf A}){\bf 1}^T\big)^{-2} \big({\bf e}_j{\bf A} \exp(t{\bf A}){\bf 1}^T\big) \big({\bf e}_j {\bf A} \exp(t{\bf A})\big)\\
    &\quad +2\big({\bf e}_j\exp(t{\bf A}){\bf 1}^T\big)^{-3}\big({\bf e}_j{\bf A}\exp(t{\bf A}){\bf 1}^T\big)^2 \big({\bf e}_j \exp(t{\bf A})\big) \\
    &\quad -\big({\bf e}_j\exp(t{\bf A}){\bf 1}^T\big)^{-2}\big({\bf e}_j{\bf A}^2\exp(t{\bf A}){\bf 1}^T\big) \big({\bf e}_j \exp(t{\bf A})\big) \\
    &\quad -\big({\bf e}_j\exp(t{\bf A}){\bf 1}^T\big)^{-2}\big({\bf e}_j{\bf A}\exp(t{\bf A}){\bf 1}^T\big) \big({\bf e}_j {\bf A} \exp(t{\bf A})\big).
\end{align*}
 \end{linenomath*}
Since 
$\exp({\bf 0}) = {\bf I}$, ${\bf e}_j{\bf 1}^T=1$, ${\bf e}_j{\bf A}{\bf 1}^T = \lambda_j$, ${\bf v}^{(j)} = {\bf e}_j{\bf A}^{[-j]}$ and
${\bf Q}^{(j)} = {\bf I}-{\bf 1}^T{\bf e}_j$,
 \begin{linenomath*}
\begin{align*} 
    \textstyle \frac{d^2}{dt^2} {\bf p}^{(j)}(t) \big|_{t = 0} 
    &= 2\lambda_j{\bf e}_j \big(\lambda_j{\bf I}-{\bf A}\big) + {\bf e}_j{\bf A}^2\big({\bf I}-{\bf1}^T{\bf e}_j\big) \nonumber \\
    &= -2\lambda_j{\bf v}^{(j)} + {\bf e}_j{\bf A}^{2} {\bf Q}^{(j)}.
\end{align*}
 \end{linenomath*}
Recalling that ${\bf A} = {\bf A}^{[-j]} + \lambda_j{\bf I}$ by \eqref{eq:A-jdef}, we can write
\begin{align} \label{eq:ejA2expr}
        {\bf e}_j{\bf A}^2 &= {\bf e}_j\big({\bf A}^{[-j]}\big)^2 + 2 \lambda_j {\bf e}_j {\bf A}^{[-j]}  + \lambda_j^2{\bf e}_j \nonumber \\
        &= {\bf v}^{(j)} {\bf A}^{[-j]} + 2 \lambda_j {\bf v}^{(j)} + \lambda_j^2{\bf e}_j.
\end{align}
Since ${\bf e}_j{\bf Q}^{(j)} = {\bf 0}$ and ${\bf v}^{(j)}{\bf Q}^{(j)}={\bf v}^{(j)}$ by \eqref{eq:ejQjzero}, it follows that
\begin{align} \label{eq:ejA2Qexpr}
    {\bf e}_j{\bf A}^2 {\bf Q}^{(j)}  &=  {\bf v}^{(j)} {\bf A}^{[-j]} {\bf Q}^{(j)} + 2 \lambda_j {\bf v}^{(j)} {\bf Q}^{(j)} =  {\bf v}^{(j)} {\bf A}^{[-j]} {\bf Q}^{(j)} + 2 \lambda_j {\bf v}^{(j)},
\end{align}
which implies
\begin{align} \label{eq:firstmomsecondderivfracresult}
    \textstyle \frac{d^2}{dt^2} {\bf p}^{(j)}(t) \big|_{t = 0} 
    = {\bf v}^{(j)} {\bf A}^{[-j]} {\bf Q}^{(j)}.
\end{align}
 It is straightforward to verify that for $i \neq j$,
  \begin{linenomath*}
    \begin{align*}
        &\textstyle \big({\bf v}^{(j)}{\bf A}^{[-j]}{\bf Q}^{(j)}\big)_i \\
        &= \textstyle \nu_{ji}(\lambda_i-\lambda_j)-\nu_{ij}\big(\sum_{k \neq j} \nu_{jk}\big)  -\nu_{ji}\big(\sum_{\ell \neq i} \nu_{i\ell}\big) + \sum_{m \neq j, m \neq i} \nu_{jm}\nu_{im}.
    \end{align*}
     \end{linenomath*}
    If $(\nu_{ik})_{k \neq i}$ and $\frac{d^2}{dt^2} {\bf p}^{(j)}(t)\big|_{t=0}$ are known, we can therefore use \eqref{eq:firstmomsecondderivfracresult} 
    to get an equation for $\lambda_i-\lambda_j$ of the form $\nu_{ji}(\lambda_i-\lambda_j) = C$ for some constant $C$.
    If $\nu_{ji} \neq 0$, we immediately obtain the value of $\lambda_i-\lambda_j$.
    If $\nu_{ji}=0$, then by our assumption of irreducibility, there exist integers $n_1,\ldots,n_k$ so that $\nu_{n_0n_1} \nu_{n_1n_2} \cdots \nu_{n_kn_{k+1}}>0$, where $n_0=j$ and $n_{k+1}=i$.
    For each $\ell=0,\ldots,k$, we can use the fact that $\nu_{{n_\ell} n_{\ell+1}}>0$ to obtain the value of $\lambda_{n_{\ell+1}}-\lambda_{n_\ell}$.
    Since $\lambda_{n_{k+1}}-\lambda_{n_0} = \sum_{\ell=0}^k (\lambda_{n_{\ell+1}}-\lambda_{n_\ell})$, we also obtain the value of $\lambda_i-\lambda_j$.
    \item We know from \eqref{eq:pfirstder} that
     \begin{linenomath*}
\begin{align*}
\begin{split}
        \textstyle \frac{d}{dt} {\bf p}^{(j)}(t) &= \big({\bf e}_j\exp(t{\bf A}){\bf 1}^T\big)^{-1} \big({\bf e}_j {\bf A} \exp(t{\bf A})\big) \\
    &\quad -\big({\bf e}_j\exp(t{\bf A}){\bf 1}^T\big)^{-2}\big({\bf e}_j{\bf A}\exp(t{\bf A}){\bf 1}^T\big) \big({\bf e}_j \exp(t{\bf A})\big).
\end{split}
\end{align*}
 \end{linenomath*}
We also know from \eqref{eq:Meanmatrixconvergence} that
\[
    \textstyle \lim_{t \to \infty} e^{-\sigma t} \exp\big(t{\bf A}\big) = {\boldsymbol\beta}^T{\boldsymbol\gamma},
\]
where $\boldsymbol\beta$ and $\boldsymbol\gamma$ are positive vectors.
It follows that as $t \to \infty$,
 \begin{linenomath*}
\begin{align*}
     \textstyle \frac{d}{dt} {\bf p}^{(j)}(t) &\to \big({\bf e}_j{\boldsymbol\beta}^T{\boldsymbol\gamma}{\bf 1}^T\big)^{-1} \big({\bf e}_j {\bf A} {\boldsymbol\beta}^T{\boldsymbol\gamma} \big) -\big({\bf e}_j{\boldsymbol\beta}^T{\boldsymbol\gamma}{\bf 1}^T\big)^{-2}\big({\bf e}_j{\bf A}{\boldsymbol\beta}^T{\boldsymbol\gamma}{\bf 1}^T\big) \big({\bf e}_j{\boldsymbol\beta}^T{\boldsymbol\gamma}\big) \\
     &= \big({\bf e}_j{\boldsymbol\beta}^T\,\overline{\boldsymbol\gamma}{\bf 1}^T\big)^{-1} \big({\bf e}_j {\bf A} {\boldsymbol\beta}^T\,\overline{\boldsymbol\gamma} \big) -\big({\bf e}_j{\boldsymbol\beta}^T\,\overline{\boldsymbol\gamma}{\bf 1}^T\big)^{-2}\big({\bf e}_j{\bf A}{\boldsymbol\beta}^T\,\overline{\boldsymbol\gamma}{\bf 1}^T\big) \big({\bf e}_j{\boldsymbol\beta}^T\,\overline{\boldsymbol\gamma}\big),
\end{align*}
 \end{linenomath*}
where $\overline{\boldsymbol\gamma}$ is the normalized version of $\boldsymbol\gamma$, see \eqref{eq:vbardef}.
Since ${\bf e}_j{\boldsymbol\beta}^T=\beta_j>0$ and $\overline{\boldsymbol\gamma}{\bf 1}^T=1$, we obtain
\begin{align} \label{eq:deriveinfty1}
     \textstyle \frac{d}{dt} {\bf p}^{(j)}(t) &\to  \beta_j^{-1} \big({\bf e}_j {\bf A} {\boldsymbol\beta}^T\,\overline{\boldsymbol\gamma}  - {\bf e}_j{\bf A}{\boldsymbol\beta}^T \overline{\boldsymbol\gamma}\,\big) = {\bf 0}.
\end{align}
On the other hand, by noting that ${\bf A}$ and $\exp(t{\bf A})$ commute, we can rewrite the expression \eqref{eq:pfirstder} for $\frac{d}{dt} {\bf p}^{(j)}(t)$ as
 \begin{linenomath*}
\begin{align*}
\begin{split}
        \textstyle \frac{d}{dt} {\bf p}^{(j)}(t) &= \big({\bf e}_j\exp(t{\bf A}){\bf 1}^T\big)^{-1} \big({\bf e}_j \exp(t{\bf A})  {\bf A}\big) \\
    &\quad -\big({\bf e}_j\exp(t{\bf A}){\bf 1}^T\big)^{-2}\big({\bf e}_j\exp(t{\bf A}){\bf A}{\bf 1}^T\big) \big({\bf e}_j \exp(t{\bf A})\big).
\end{split}
\end{align*}
 \end{linenomath*}
Since ${\bf A}{\bf 1}^T = {\boldsymbol\lambda}^T$, ${\bf A} = {\bf V} + {\rm diag}(\boldsymbol\lambda)$ by \eqref{eq:Vdef}, $\overline{\boldsymbol\gamma} \, {\rm diag}(\boldsymbol\lambda) = {\boldsymbol\lambda} \, {\rm diag}\big(\overline{\boldsymbol\gamma}\big)$
and $\overline{\boldsymbol\gamma} {\boldsymbol\lambda}^T = {\boldsymbol\lambda}\, \overline{\boldsymbol\gamma}^T$, we get as $t \to \infty$,
\begin{align} \label{eq:deriveinfty2}
            \textstyle \frac{d}{dt} {\bf p}^{(j)}(t) &\to \overline{\boldsymbol\gamma} {\bf A} -  \overline{\boldsymbol\gamma} {\bf A}{\bf 1}^T \, \overline{\boldsymbol\gamma} = \overline{\boldsymbol\gamma} \big({\bf A} - {\boldsymbol\lambda}^T\overline{\boldsymbol\gamma}\,\big)
            = \overline{\boldsymbol\gamma} \, {\bf V} 
            + {\boldsymbol\lambda} \, {\rm diag} \big(\overline{\boldsymbol\gamma}\big) 
            - {\boldsymbol\lambda}\, \overline{\boldsymbol\gamma}^T\, \overline{\boldsymbol\gamma}.
\end{align}
Combining \eqref{eq:deriveinfty1} and \eqref{eq:deriveinfty2}, we obtain the following linear system for $\boldsymbol\lambda$:
\[
\boldsymbol\lambda \big({\rm diag} \big(\overline{\boldsymbol\gamma}\big) - \overline{\boldsymbol\gamma}^T\, \overline{\boldsymbol\gamma}\,\big) = -\overline{\boldsymbol\gamma} \, {\bf V}.
\]
It is straightforward to verify that this system is solved by
\[
    \boldsymbol\lambda = {\bf a} + x {\bf 1}, \quad x \in \mathbb{R},
\]
for some vector ${\bf a}$,
which can be used to extract $\boldsymbol\lambda^{[-1]}$.
\end{enumerate}
\item By the definition of ${\bf S}^{(j)}(t)$ in \eqref{eq:QalphaSalphadef},
\begin{align} \label{eq:deriv}
    \textstyle \frac{d}{dt} {\bf S}^{(j)}(t) &= \textstyle \frac{d}{dt} \big({\bf e}_j{\bf M}(t){\bf 1}^T\big)^{-2}\big({\bf Q}^{(j)}(t)\big)^T{\boldsymbol\Sigma}^{(j)}(t){\bf Q}^{(j)}(t) \nonumber \\
    &\quad \textstyle + \big({\bf e}_j{\bf M}(t){\bf 1}^T\big)^{-2} \frac{d}{dt} \big({\bf Q}^{(j)}(t)\big)^T{\boldsymbol\Sigma}^{(j)}(t){\bf Q}^{(j)}(t) \nonumber \\
    &\quad \textstyle + \big({\bf e}_j{\bf M}(t){\bf 1}^T\big)^{-2}\big({\bf Q}^{(j)}(t)\big)^T \frac{d}{dt} {\boldsymbol\Sigma}^{(j)}(t){\bf Q}^{(j)}(t) \nonumber \\
    &\quad \textstyle + \big({\bf e}_j{\bf M}(t){\bf 1}^T\big)^{-2}\big({\bf Q}^{(j)}(t)\big)^T{\boldsymbol\Sigma}^{(j)}(t)\frac{d}{dt}{\bf Q}^{(j)}(t).
\end{align}
Since ${\boldsymbol\Sigma}^{(j)}(0) = {\bf 0}$ and ${\bf e}_j{\bf M}(0){\bf 1}^T=1$, we obtain
\[
    \textstyle \frac{d}{dt} {\bf S}^{(j)}(t)\big|_{t = 0} = \textstyle 
    \big({\bf Q}^{(j)}\big)^T \big(\frac{d}{dt} {\boldsymbol\Sigma}^{(j)}(t)\big|_{t = 0}\big){\bf Q}^{(j)}.
\]
From \eqref{eq:ddtSigma} in the proof of Proposition \ref{cor:idcellnum}, we know that
\begin{align} \label{eq:Sigmaderivexpr}
        \textstyle \frac{d}{dt} {\boldsymbol\Sigma}^{(j)}(t)\big|_{t = 0} &= 2b_j{\bf e}_j^T{\bf e}_j + {\rm diag}\big({\bf e}_j{\bf A}\big) - \big({\bf e}_j^T{\bf e}_j{\bf A}\big)^T - {\bf e}_j^T{\bf e}_j{\bf A} \nonumber \\
    &= {\rm diag}\big({\bf e}_j{\bf A}^{[-j]}\big) - \big({\bf e}_j^T{\bf e}_j{\bf A}\big)^T - {\bf e}_j^T{\bf e}_j{\bf A} + (2b_j+\lambda_j){\bf e}_j^T{\bf e}_j,
\end{align}
where in the second step, we write ${\bf A} = {\bf A}^{[-j]} + \lambda_j{\bf I}$.
Since ${\bf e}_j{\bf Q}^{(j)} = {\bf 0}$ 
and ${\bf e}_j {\bf A}^{[-j]} = {\bf v}^{(j)}$,
we obtain
\begin{align} \label{eq:secomomderivresult}
     &\textstyle 
     \frac{d}{dt} {\bf S}^{(j)}(t)\big|_{t = 0} 
     = 
     \big({\bf Q}^{(j)}\big)^T {\rm diag}\big({\bf e}_j{\bf A}^{[-j]}\big) {\bf Q}^{(j)}
     = \big({\bf Q}^{(j)}\big)^T {\rm diag}\big({\bf v}^{(j)}\big) {\bf Q}^{(j)}.
\end{align}
It is straightforward to verify that the $(j,k)$-th coordinate of $\big({\bf Q}^{(j)}\big)^T {\rm diag}\big({\bf v}^{(j)}\big) {\bf Q}^{(j)}$
is $-\nu_{jk}$.
Thus, knowledge of the switching rates $(\nu_{ik})_{k \neq i}$ follows immediately from knowledge of $\frac{d}{dt} {\bf S}^{(j)}(t)\big|_{t = 0}$ for $j=1,\ldots,K$, but no other parameters can be extracted.
\vspace*{12pt}
\qedhere
\end{enumerate}
\end{proof}

\section{Implementation in MATLAB} \label{app:implementation}

In this section, we give details on how our estimation framework is implemented in MATLAB.

\subsection{Cell number data}

The first step in the implementation for cell number data is to compute simple parameter estimates for the switching rates $(\nu_{ik})_{k \neq i}$ and the net birth rates $\boldsymbol\lambda$ based on a deterministic population model.
This model is obtained by ignoring the stochastic terms in the statistical model \eqref{eq:statisticalmodelnum}, i.e.~by equating the data vector ${\bf n}_{i,\ell,r}$ with the mean prediction of \eqref{eq:statisticalmodelnum}:
\begin{align} \label{eq:deterministicapproxnum}
    {\bf n}_{i,\ell,r} = N_i {\bf m}^{{\bf f}_i}(t_\ell) = {\bf n}_i {\bf M}(t_\ell).
\end{align}
Let ${\bf N}$ be the $I \times K$ matrix with the initial conditions ${\bf n}_i$ as row vectors, and let ${\bf N}_{\ell,r}$ be the $I \times K$ matrix with the data vectors ${\bf n}_{i,\ell,r}$ as row vectors.
We can then write \eqref{eq:deterministicapproxnum} in matrix form as
\begin{align} \label{eq:deterministicapproxnum2}
    {\bf N}_{\ell,r} = {\bf N} {\bf M}(t_\ell) = {\bf N} \exp(t_\ell{\bf A}).
\end{align}
Assuming ${\bf N}$ has rank $K$,
we can solve for ${\bf A}$ in \eqref{eq:deterministicapproxnum2} by first multiplying both sides by ${\bf N}^T$, then multiplying both sides by the inverse of ${\bf N}^T{\bf N}$, and finally taking a matrix logarithm.
We can  thus obtain an estimate for the infinitesimal generator ${\bf A}$,
\[
    {\bf A}_{\ell,r}^\ast := (1/t_\ell) \log\big(\big({\bf N}^T{\bf N}\big)^{-1} {\bf N}^T\,{\bf N}_{\ell,r}\big).
\]
We then compute a final estimate ${\bf A}^\ast$ by averaging across time points and replicates:
\begin{align} \label{eq:simpleestimatenum}
    \textstyle {\bf A}^\ast := (1/(L R)) \sum_{\ell=1}^{L} \sum_{r=1}^{R} {\bf A}_{\ell,r}^\ast.
\end{align}
From ${\bf A}^\ast$, we can obtain estimates of the switching rates $(\nu_{ik})_{k \neq i}$ and the net birth rates $\boldsymbol\lambda$.

As indicated in Appendix \ref{sec:extensions},
we implement the following likelihood function in our codes:
\begin{linenomath*}
    \begin{align*}
    &{\cal L}_{\rm num}\big(
    {\boldsymbol\theta}_{\rm num}
    \big|({\bf n}_{i,\ell,r})_{i,\ell,r}\big) \nonumber \\
    &= \textstyle \prod_{i=1}^I \prod_{\ell=1}^{L} \prod_{r=1}^{R} \! \Big((2\pi)^{K} {\rm det}\big({\bf C}_i^T \big( N_i{\boldsymbol\Sigma}^{{\bf f}_i}(t_\ell)+{\bf E}_{i,\ell}^{\rm num}\big) {\bf C}_i \big)\Big)^{-1/2} \nonumber \\
    &\quad \textstyle \cdot\exp\big(\!-\!\frac12 \big({\bf n}_{i,\ell,r}-N_i{\bf m}^{{\bf f}_i}(t_\ell)\big) {\bf C}_i  \big({\bf C}_i^T \big( N_i{\boldsymbol\Sigma}^{{\bf f}_i}(t_\ell)+{\bf E}_{i,\ell}^{\rm num}\big) {\bf C}_i \big)^{-1} {\bf C}_i^T \big({\bf n}_{i,\ell,r}-N_i{\bf m}^{{\bf f}_i}(t_\ell)\big)^T\big).
\end{align*}
\end{linenomath*}
For each $i = 1,\ldots,I$, ${\bf C}_i$ is a $K \times J_i$ matrix for some $1 \leq J_i \leq K$, which 
can be used to reduce the dimension of the data vector ${\bf n}_{i,\ell,r}$ when necessary.
This option can e.g.~be useful 
for models with reducible switching dynamics, see Appendix \ref{sec:extensions}.

From the above likelihood function, we compute a negative double log-likelihood as in \eqref{eq:loglikelihood}, and solve the MLE problem \eqref{eq:mleestimator} using the sequential quadratic programming (sqp) solver in MATLAB.
For the optimization, one must supply an initial guess $\boldsymbol\theta_{{\rm num}}^{(0)}$ for the parameter vector $\boldsymbol\theta_{\rm num}$,
and a set of feasible parameters $\boldsymbol\Theta_{\rm num}$ of the form
\[
    \boldsymbol\Theta_{\rm num} = \{\boldsymbol\theta_{\rm num}: {\bf l} \leq \boldsymbol\theta_{\rm num} \leq {\bf u}, {\bf G} \, \boldsymbol\theta_{\rm num} \leq {\bf h}, {\bf G}_{\rm eq} \,\boldsymbol\theta_{\rm num} = {\bf h}_{\rm eq}\}.
\]
By default, we assume lower bounds of {\bf 0} for the switching rates $(\nu_{ik})_{k \neq i}$ and the birth rates ${\bf b}$, and we impose the inequality constraint $\boldsymbol\lambda \leq {\bf b}$.
The user is expected to provide lower bounds for the net birth rates $\boldsymbol\lambda$
and upper bounds for all parameters, and they have the option to provide further inequality or equality constraints as necessary.
This provides the opportunity to impose constraints such as $\lambda_1=\lambda_2$ (Section \ref{sec:realdata}) or $\nu_{13}=\nu_{31}=\nu_{32}=0$ (Appendix \ref{sec:extensions}).

For the initial guess $\boldsymbol\theta_{{\rm num}}^{(0)}$,
we use the simple estimates for $(\nu_{ik})_{k \neq i}$ and $\boldsymbol\lambda$
computed from \eqref{eq:simpleestimatenum}.
An initial guess for the birth rate $b_i$ is generated as
$|\lambda_i|/U$, where $U$ is uniformly distributed between 0 and 1.
The idea is that if $\lambda_i > 0$,
then in the absence of phenotypic switching,
the survival probability of a single-cell derived clone of type $i$ 
is $q_i = \lambda_i/b_i$ \cite{durrett2015branching}.
Since we do not assume any information on $q_i$, we sample it uniformly between 0 and 1,
and then use the initial guess for $\lambda_i$ to compute an initial guess for $b_i$.

If data on the number of dead cells at each time point is available, the initial guesses for the birth rates can be improved as follows.
As before, let $n_{ij}$ be the number of starting cells of type-$j$ under the $i$-th initial condition. In the absence of phenotypic switching, the expected number of type-$j$ cells at time $t$ under the $i$-th initial condition is given by $n_{ij} \exp(\lambda_j t)$. If we assume that type-$j$ cells grow deterministically according to this function, the number of dead cells of type-$j$ that accumulate up until the first experimental timepoint $t_1$ is given by
\[
    \textstyle d_j \int_0^{t_1} n_{ij} \exp(\lambda_j t)dt  = d_j \lambda_j^{-1} n_{ij} \big(\exp(\lambda_j t_1) - 1\big).
\]
Set $D_{ij} := \lambda_j^{-1} n_{ij} \big(\exp(\lambda_j t_1) - 1\big)$ and let ${\bf D} = (D_{ij})$ denote the corresponding $I \times K$ matrix.
Also, let ${\bf c}$ denote the $1 \times I$ vector of the experimentally measured number of dead cells at time $t_1$, averaged across the $R$ experimental replicates.
We should then have
\[
    {\bf d} {\bf D}^T = {\bf c}.
\]
Assuming ${\bf D}$ has rank $K$, we can solve this equation for ${\bf d}$ as follows:
\[
    {\bf d} = {\bf c} {\bf D} \big({\bf D}^T{\bf D}\big)^{-1},
\]
which gives an estimate for the vector of death rates ${\bf d}$. An estimate for the birth rates ${\bf b}$ can then be computed as ${\bf b} = {\boldsymbol\lambda} + {\bf d}$.

In addition to being used to initialize the optimization,
the initial guess $\boldsymbol\theta_{\rm num}^{(0)}$ is used to estimate the
relative scales of the parameters $(\nu_{ik})_{k \neq i}$, $\boldsymbol\lambda$ and ${\bf b}$.
In particular, for the $i$-th coordinate of the initial guess,
we define the corresponding scale variable
\[
    s_{i}^{(0)} := 10^{\lfloor \log_{10}|\theta_{{\rm num},i}^{(0)}|\rfloor},
\]
with $s_i^{(0)} := 1$ if $\theta_{{\rm num},i}^{(0)} = 0$. For example, if the initial guesses are ${\bf b}^{(0)} = (1.5,1.2)$ for the birth rates, $\boldsymbol\lambda^{(0)} = (0.3,0.4)$ for the net birth rates, and $(\nu_{12}^{(0)},\nu_{21}^{(0)}) = (0.05,0.002)$ for the switching rates, the corresponding scale variables are $(1,1)$, $(0.1,0.1)$ and $(0.01,0.001)$, respectively.
For a given parameter vector $\boldsymbol\theta_{\rm num}$,
we define the transformed vector
\[
    \widetilde{\boldsymbol\theta}_{\rm num} := {\boldsymbol\theta}_{\rm num} \oslash {\bf s}^{(0)},
\]
where $\oslash$ denotes elementwise division.
For the initial guesses ${\bf b}^{(0)} = (1.5,1.2)$, $\boldsymbol\lambda^{(0)} = (0.3,0.4)$ and $(\nu_{12}^{(0)},\nu_{21}^{(0)}) = (0.05,0.002)$,
the corresponding transformed values are $\widetilde{\bf b}^{(0)} = (1.5,1.2)$, $\widetilde{\boldsymbol\lambda}^{(0)} = (3,4)$ and $\widetilde{\boldsymbol\lambda}^{(0)} = (5,2)$.
With this transformation, all nonzero parameters take values in $[1,10]$.
When we solve the MLE problem \eqref{eq:mleestimator}, we treat $ \widetilde{\boldsymbol\theta}_{\rm num}$ as the parameter vector 
instead of ${\boldsymbol\theta}_{\rm num}$, 
and solve 
\begin{align} \label{eq:transformedproblemnum}
    \textstyle \min_{\widetilde{\boldsymbol\theta}_{\rm num} \in \widetilde{\boldsymbol\Theta}_{\rm num}} \, 
    l_{\rm num}\big(\widetilde{\boldsymbol\theta}_{\rm num} \odot {\bf s}^{(0)}\big),
\end{align}
where $\widetilde{\boldsymbol\Theta}_{\rm num}$ is the transformed set of feasible parameters.
The parameter scaling is applied to ensure that all model parameters
are of a similar 
magnitude in the optimization.

In most cases, we have found it sufficient to solve the optimization problem \eqref{eq:transformedproblemnum} once.
However, in our codes, we provide an option to solve the problem
multiple  times, using
(i) user-supplied initial guesses,
(ii) initial guesses based on the simple estimates from \eqref{eq:simpleestimatenum}, with new birth rates selected randomly each time,
or (iii) randomly sampled initial guesses, using the parameter generation procedure described in Appendix \ref{app:numericalexp} below.

The optimization problems \eqref{eq:cioptproblem} for the endpoints of the confidence intervals are solved in a similar way, except the initial guess is taken to be the maximum likelihood estimate.

\begin{figure}
    \centering
    \includegraphics{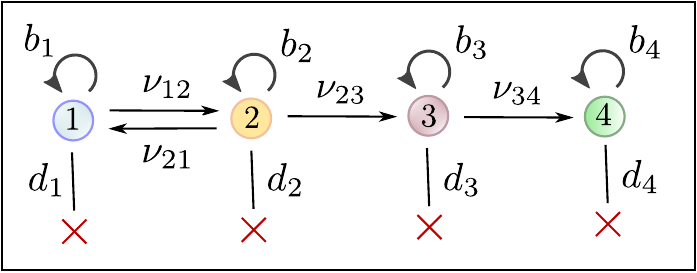}
    \caption[An example of a four-type switching model where the likelihood function \eqref{eq:likelihoodfrac} for cell fraction data from the main text must be modified to avoid degeneracy issues.]{An example of a four-type switching model where the likelihood function \eqref{eq:likelihoodfrac} for cell fraction data from the main text must be modified to avoid degeneracy issues.
    This model structure can e.g.~arise in the context of epigenetically-driven drug resistance in cancer, where drug-sensitive (type-0) cells can acquire transient resistance (type-1), which then evolves gradually to stable resistance (type-4) in two steps \cite{gunnarsson2020understanding}.
    }
    \label{fig:extension_fig_2}
\end{figure}

\subsection{Cell fraction data}

The implementation for cell fraction data is similar with the following modifications.
First of all, we parametrize the model in terms of the death rates ${\bf d}$, the net birth rate $\lambda_1$ and the net birth rate differences $\boldsymbol\lambda^{[-1]}$, instead of the birth rates ${\bf b}$ and net birth rates $\boldsymbol\lambda$.
Second, the initial guess for the MLE problem \eqref{eq:mleestimator2} is based on solving the following least squares problem, which minimizes the sum of squared errors between the mean prediction
of the statistical model \eqref{eq:statisticalmodelfrac} and the data:
\begin{align} \label{eq:simpleestimatefrac}
    & \textstyle \big\{\widehat{\boldsymbol\lambda}^{[-1]},(\widehat{\nu}_{ik})_{k \neq i}\textstyle\} = {\rm argmin }_{\boldsymbol\lambda^{[-1]},(\nu_{ik})_{k \neq i}} \sum_{i=1}^I \sum_{\ell=1}^{L} \sum_{r=1}^{R} \big\|{\bf f}_{i,\ell,r} - {\bf p}^{{\bf f}_i}(t_\ell)\big\|^2.
\end{align}
Note that this is 
a continuous-time version of the TRANSCOMPP problem \eqref{eq:TRANSCOMPP}.
When solving \eqref{eq:simpleestimatefrac}, we need to supply an initial guess.
If experiments are conducted from isolated initial conditions, we compute initial guesses for the switching rates $(\nu_{ik})_{k \neq i}$ based on part (1) of Proposition \ref{cor:idcellfrac1}, which shows how $(\nu_{ik})_{k \neq i}$ can be estimated from the slopes of the mean functions ${\bf p}^{(j)}(t)$ at time zero.
We approximate the slopes of ${\bf p}^{(j)}(t)$ at time zero using experimentally observed cell fractions at the first time point.
The initial guesses for the remaining parameters are set to 0.
If experiments are not conducted from isolated initial conditions, we randomly sample initial guesses as described in Appendix \ref{app:numericalexp} below.
The simple problem \eqref{eq:simpleestimatefrac} returns estimates for
$(\nu_{ik})_{k \neq i}$ and $\boldsymbol\lambda^{[-1]}$, which we supply as initial guesses
to 
\eqref{eq:mleestimator2}.

In our codes, we implement the following likelihood function for cell fraction data:
\begin{linenomath*}
    \begin{align*}
    &{\cal L}_{\rm frac}\big(\boldsymbol\theta_{\rm frac}\big|({\bf f}_{i,\ell,r})_{i,\ell,r}\big) \nonumber \\
    &= \textstyle \prod_{i=1}^I \prod_{\ell=1}^{L} \prod_{r=1}^{R} \! \Big((2\pi)^{K-1} {\rm det}\big({\bf C}_i^T {\bf B}^T \big( N_i^{-1}{\bf S}^{{\bf f}_i}(t_\ell)+{\bf E}_{i,\ell}^{\rm frac}\big) {\bf B} {\bf C}_i\big)\Big)^{-1/2} \nonumber \\
    &\quad \textstyle \cdot\exp\Big(\!-\!\frac12 \big({\bf f}_{i,\ell,r}-{\bf p}^{{\bf f}_i}(t_\ell) \big) \,{\bf B} {\bf C}_i\, \textstyle \big({\bf C}_i^T {\bf B}^T\big( N_i^{-1} {\bf S}^{{\bf f}_i}(t_\ell)+{\bf E}_{i,\ell}^{\rm frac}\big)\,{\bf B} {\bf C}_i\big)^{-1} 
    \\
    &\quad \quad\quad \quad\; 
    \, {\bf C}_i^T {\bf B}^T \, \big({\bf f}_{i,\ell,r}-{\bf p}^{{\bf f}_i}(t_\ell)\big)^T\Big).
\end{align*}
\end{linenomath*}
Recall from \eqref{eq:likelihoodfrac} that the matrix ${\bf B}$ is applied to reduce the data vector ${\bf f}_{i,\ell,r}$ to a $(K-1)$-dimensional vector.
To accommodate reducible switching dynamics, 
the user is allowed to implement a further reduction in the data by specifying a $(K-1) \times J_i$ matrix ${\bf C}_i$
for each initial condition $i$.
This can for example be useful for the four-type model ($K=4$) displayed in Figure \ref{fig:extension_fig_2},
in which case we would take $I=3$,
${\bf C}_1 = {\bf C}_2 = {\bf I}$ and ${\bf C}_3 = {\bf e}_3^T$,
and we would restrict the set of feasible parameters $\boldsymbol\Theta_{\rm frac}$ so that $\nu_{13} = \nu_{14} = \nu_{24} =  \nu_{31} = \nu_{32} = \nu_{41} = \nu_{42} = \nu_{43} = 0$.
Note that here, ${\bf I}$ refers to the $(K-1) \times (K-1) = 3 \times 3$ identity matrix.

\section{Additional numerical results} \label{app:results}

This section contains additional numerical results to those discussed in Section \ref{sec:numerical} of the main text.

\begin{figure}
    \centering
    \includegraphics[scale=0.8]{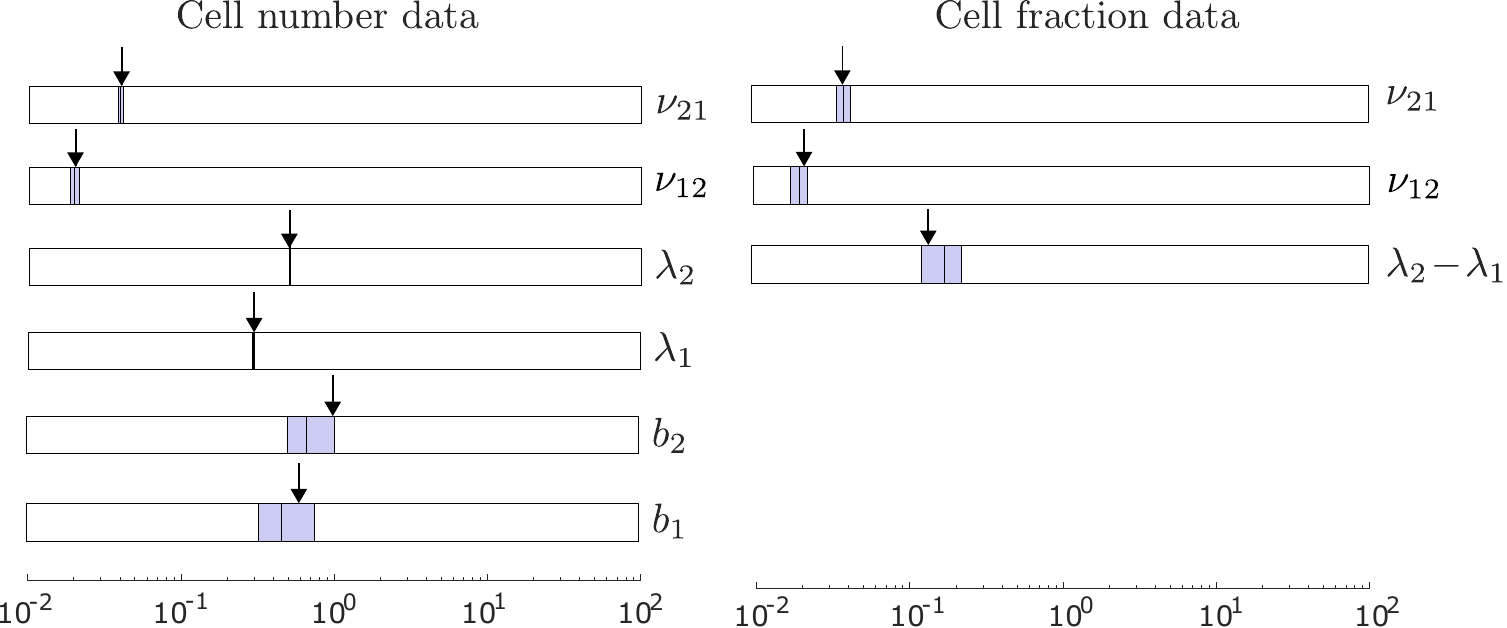}
    \caption[Graphical depiction of the output of our estimation framework.]{Graphical depiction of the output of our estimation framework.
    We first generated artificial cell-number and cell-fraction data by simulating the branching process model of Section \ref{sec:model} for $b_1=0.6$, $d_1 = 0.3$, $b_2 = 1.0$, $d_2=0.5$, $\nu_{12} = 0.02$, $\nu_{21} = 0.04$ and $N_1=N_2=1,000$.
    Using this data, we computed maximum likelihood estimates and likelihood-based 95\% confidence intervals (CIs) for the model parameters.
    For each parameter, the shaded region indicates the CI,
    the vertical bar inside the interval indicates the MLE estimate,
    and the arrow points to the true value of the parameter.
    }
    \label{fig:itworks}
\end{figure}

\subsection{Illustrative example} \label{sec:partexample}

For illustrative purposes,
we 
show here 
a graphical depiction of 
the output of our estimation framework
for a single dataset.
We generated artificial 
cell number and cell fraction data 
by performing a stochastic
simulation of the branching process model from Section \ref{sec:model}.
We then used the data to compute MLE estimates and confidence intervals for the model parameters.
The data was generated assuming $K=2$ cell types, isolated initial conditions,
$L=6$ time points, and $R=3$ replicates.
Estimation results are shown in Figure \ref{fig:itworks}.

Note first the difference in scale between the switching rates
and the rates involving cell division and death.
This is typically the case, 
since epigenetic modifications
can generally be retained for $10$--$10^5$ cell divisions
\cite{niepel2009non,brown2014poised}.
Also note that 
all model parameters are estimated more accurately
for cell number data than cell fraction data,
in that their confidence intervals
are narrower for cell number data.
Otherwise, the relative accuracy with which different model parameters
can be estimated is in line with our identifiability analysis in Section \ref{sec:identifiability}.

\begin{figure}[!t]
    \centering
    \includegraphics[scale=1]{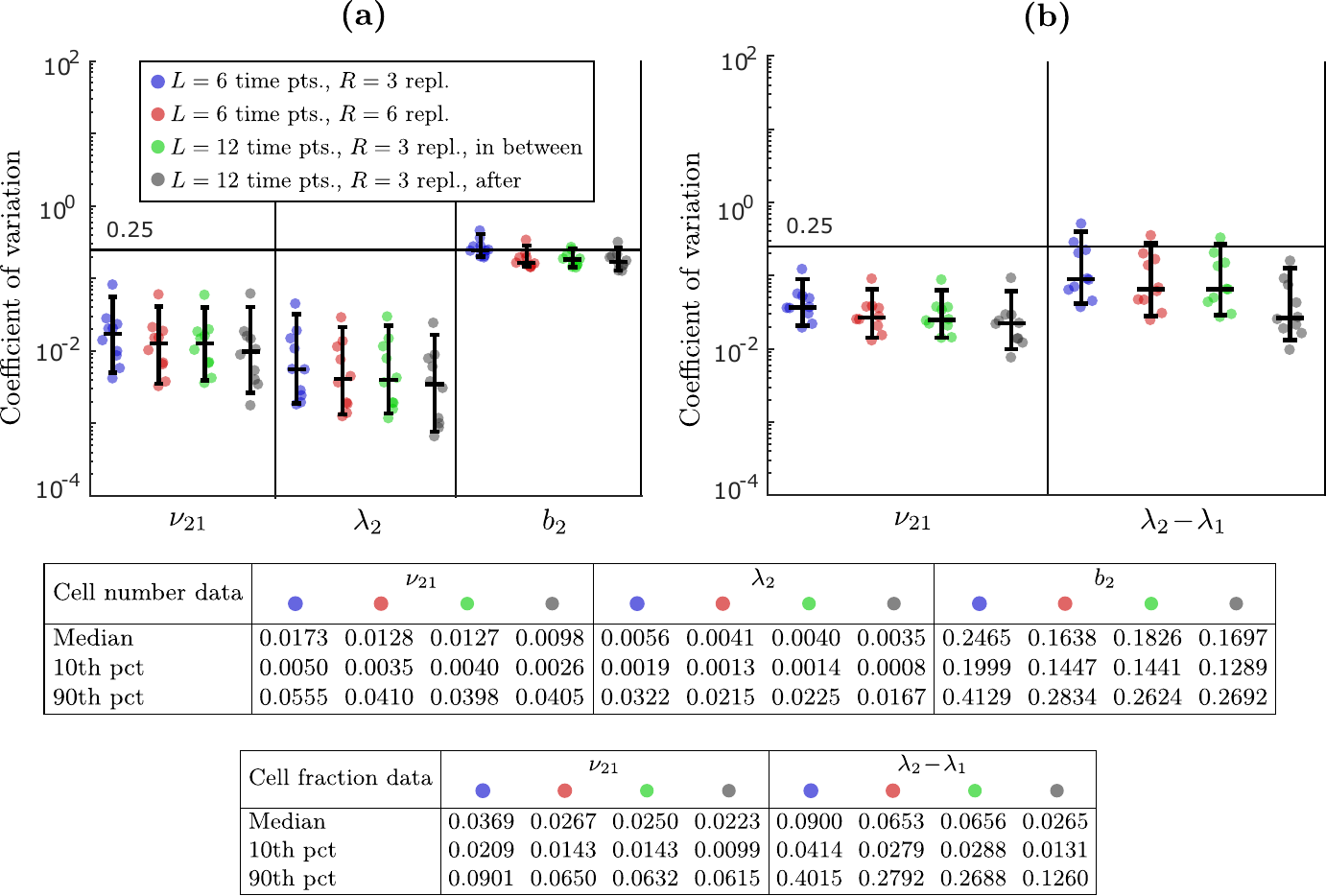}       
    \caption[Comparison of estimation error for different experimental designs
    when the number of data points is doubled.]{
    Comparison of estimation error for different experimental designs
    when the number of data points is doubled.
    We generated 10 parameter regimes and 100 datasets for each regime.
    The blue dots represent estimation from datasets with $L=6$ time points and $R=3$ replicates.
    The red dots represent estimation from $L=6$ time points and $R=6$ replicates.
    The green and grey dots represent estimation from $L=12$ time points and $R=3$ replicates,
    where the extra time points are added in between and after the previous time points, respectively.
    Panel {\bf (a)} shows estimation from cell number data and panel {\bf (b)} shows estimation from cell fraction data. 
    }
    \label{fig:expdesign}
\end{figure}

\subsection{Experimental design: Adding replicates vs.~adding time points} \label{sec:expdesign}

In this section, we discuss how our framework can be used to evaluate to what extent additional data can improve parameter estimates
and to identify experimental designs that best accomplish this goal.
To illustrate this point, we compared the effect of 
(i) doubling the number of replicates from $R=3$ to $R=6$ (design 1),
(ii) doubling the number of time points from $T=6$ to $T=12$,
adding time points in between the previous time points (design 2),
and
(iii) doubling the number of time points,
adding time points after the previous points (design 3) (Appendix \ref{app:numericalexp}).
We generated 10 parameter regimes
and 100 datasets for each regime.
The results are shown in Figure \ref{fig:expdesign}.

For cell number data, the median CV for the switching rate $\nu_{21}$
and the net birth rate $\lambda_2$
reduces by 26\% and 27\%, respectively,
when the number of replicates is doubled (design 1)
(Fig.~\ref{fig:expdesign}a).
This is consistent with the fact that 
the standard deviation of an MLE estimator
can be expected to decrease with $1/\sqrt{n}$,
where $n$ is the number of datapoints ($1-1/\sqrt{2} = 0.29$) \cite{casella2021statistical}.
Adding data from time points in between the previous time points (design 2) has
a similar effect on the median CV.
However, adding time points after the previous points (design 3)
reduces the median CV of $\nu_{21}$ and $\lambda_2$ by 23\% and 16\%, respectively, over adding replicates (design 1).
We also note that the 10th percentile of the CV for $\nu_{21}$ and $\lambda_2$ reduces by 26\% and 42\%, respectively, between design 1 and design 3,
which indicates that the degree of improvement
between design 1 and design 3 depends very much on the parameter regime.

For cell fraction data, the relative attractiveness of the three experimental designs is similar (Fig.~\ref{fig:expdesign}b).
However, in this case, 
the estimate for the net birth rate difference $\lambda_2-\lambda_1$
benefits significantly more from using design 3 than the estimate for the switching rate $\nu_{21}$.
For example, the median CV for $\nu_{21}$ reduces by 16\% and the 10th percentile by 30\% between design 1 and design 3,
while the analogous reduction for $\lambda_2-\lambda_1$ is 59\% and 53\%, respectively.

In our structural identifiability analysis for cell fraction data (Section \ref{sec:identifiabilitycellfraction}),
we observed that it is more difficult to estimate $\lambda_2-\lambda_1$ than $\nu_{21}$ from the initial population dynamics,
and that $\lambda_2-\lambda_1$ can be identified from the equilibrium proportions
$\overline{\boldsymbol\gamma}$ if the switching rates $(\nu_{ik})_{k \neq i}$ are known.
The fact that 
adding more information
on the long-run behavior of the population benefits the estimation of $\lambda_2-\lambda_1$
more than $\nu_{21}$ is consistent with these insights.
Of course, the results of Section \ref{sec:numericallargescale} indicate that the estimation of $\lambda_2-\lambda_1$
can be improved even 
further
by using cell number data as opposed to cell fraction data.

In general, Sections \ref{sec:numericallargescale} and \ref{sec:expdesign} show how our framework can be used to evaluate
the estimation accuracy that can be achieved
by different experimental designs,
depending e.g.~on what data is collected, when it is collected, how many replicates are performed, etc.

\section{Generation of artificial data} \label{app:numericalexp}

Here, we discuss how the artificial data was generated for the numerical experiments in Section \ref{sec:numerical}.
First, to generate each parameter regime,
we sampled the birth rates ${\bf b}$ and death rates ${\bf d}$
uniformly at random on $(0,1)$, with the following caveats:
The birth rates ${\bf b}$ and net birth rates $\boldsymbol\lambda$ were required to be larger than 0.01 in absolute value, and at least one of the net birth rates $\lambda_1,\lambda_2$ was required to be positive.
Each switching rate $\nu_{ij}$ was sampled as $10^{-3+2U}$, where $U$ is uniform between 0 and 1, meaning that it was sampled log-uniformly between $10^{-3}$ and $10^{-1}$.
The starting number of cells $N_i$ was chosen as $N_i = 10^{-3}$, $N_i = 10^{-4}$ or $N_i = 10^{-5}$ for $i=1,2$ based on the order of magnitude of the smallest switching rate.
The experimental time points were selected as $t=1,\ldots,6$.

In Section \ref{sec:expdesign}, where the number of time points was doubled, the time points were taken as either $t=0.5,1,1.5,2,\ldots,6$ or $t=1,2,3,\ldots,12$,
depending on whether the new time points were added in between or after the previous time points.

Once the parameters were set, we performed stochastic simulations of the model in Section \ref{sec:model} to obtain the artificial datasets.
The parameter regimes used to perform the simulations are available
in the Github repository for the paper 
(\url{https://github.com/egunnars/phenotypic_switching_inference/}).
The background MATLAB codes used to generate the parameter regimes and the artificial datasets, and to perform estimation on the artificial datasets, 
are also available in the same repository.

\section{AIC and BIC} \label{app:AIC}

To evaluate model fit relative to model complexity in Section \ref{sec:realdata},
we use the Akaike Information Criterion (AIC) and the Bayesian Information Criterion (BIC).
For a statistical model with parameters $\boldsymbol\theta$ and negative double log-likelihood $l(\boldsymbol\theta)$, 
the AIC and BIC are given by
\begin{linenomath*}
\begin{align*}
& {\rm AIC} = l\big(\widehat{\boldsymbol\theta}\big) + 2p, \\
    & {\rm BIC} = l\big(\widehat{\boldsymbol\theta}\big) + p \log(n),
\end{align*}
\end{linenomath*}
where $\widehat{\boldsymbol\theta}$ is the MLE estimate,
$p$ is the number of parameters in the statistical model, and $n$ is the number of datapoints.
When comparing two models, the model with the lower AIC or BIC is preferred, depending on which criterion is used.
The BIC criterion generally favors simpler models, i.e.~models with fewer parameters, to a greater extent than the AIC criterion. \\

\noindent {\bf Competing Interests Statement.} The authors have no competing interests to declare. \\

\noindent {\bf Acknowledgments.}
EBG and JF were supported in part by NIH grant R01CA241137. EBG and KL were supported in part by NSF grant CMMI-1552764. JF was supported in part by NSF grants DMS-1349724 and DMS-2052465. JF and KL were supported in part by the Research Council of Norway R\&D Grant 309273. EBG was supported in part by the Norwegian Centennial Chair grant and the Doctoral Dissertation Fellowship from the University of Minnesota.

\newpage

\bibliography{epi}
\bibliographystyle{unsrt}

\end{document}